\documentclass[aps,prd,tighten,11pt,showpacs,a4,nofootinbib]{revtex4}

\usepackage[utf8]{inputenc}
 
\usepackage{graphicx}
\usepackage{bm}
\usepackage{amsmath,amssymb}
\usepackage{latexsym}
\usepackage{color}
\usepackage{tabularx}
\usepackage{upgreek}
\usepackage{mathtools}

\newcommand{\pp}{\phantom}

\newcommand{\al}{\alpha}

\newcommand{\ga}{\gamma}

\def\na{\nabla}
\newcommand{\de}{\delta}

\begin{document}

\title{Quantum fields in Bianchi type I spacetimes. The Kasner metrc.}

\author{Jerzy Matyjasek}

\affiliation{Institute of Physics,
Maria Curie-Sk\l odowska University\\
pl. Marii Curie-Sk\l odowskiej 1,
20-031 Lublin, Poland}

\begin{abstract}
Vacuum polarization of the quantized massive fields in Bianchi type I  spacetime 
is investigated from the point of view of the adiabatic approximation and 
the Schwinger-DeWitt method. It is shown that both approaches give the same 
results that can be used in construction of the trace of the stress-energy 
tensor of the conformally coupled fields. The stress-energy tensor is 
calculated in the Bianchi type I spacetime  and the back reaction of the quantized 
fields upon the Kasner geometry is studied. A special emphasis is put on 
the problem of isotropization, studied with the aid of the directional Hubble 
parameters. Similarities with the quantum corrected interior of the Schwarzschild black hole 
is briefly discussed. 
\end{abstract}
\maketitle

\section{Introduction}

\begin{figure} 
\centering
\includegraphics[width=11cm]{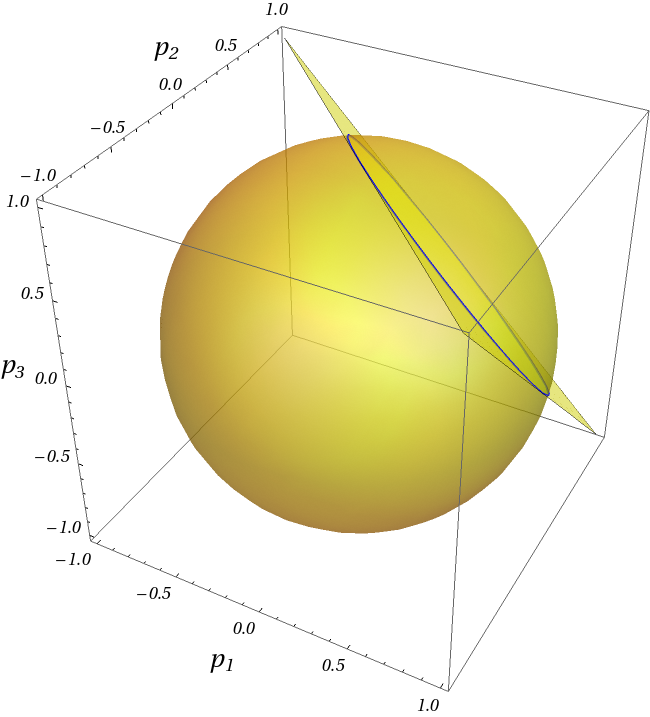}
\caption{ The Kasner sphere and the Kasner plane in the parameter space.  }
\label{fK3}
\end{figure}

\begin{figure} 
\centering
\includegraphics[width=11cm]{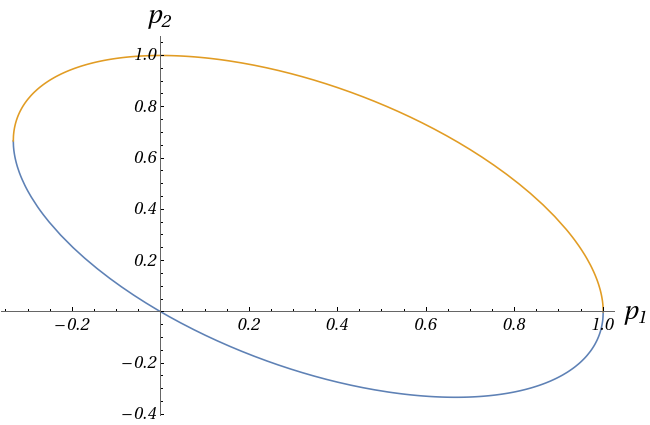}
\caption{ Two branches of the  allowable parameters in the $(p_{1}, p_{2})$ space.
The parameter $p_{3}$ can be obtained from Eq.~\ref{eq_p3}. The branch points represent
degenerate configurations $(-1/3,2/3,2/3)$ and $(1,0,0).$}
\label{fK2}
\end{figure}

In this paper we shall consider quantized massive  fields 
in homogeneous anisotropic cosmological Bianchi type I models
described by the line element
\begin{equation}
 ds^{2} = -dt^{2} + a^{2}(t) dx^{2} + b^{2}(t)dy^{2} + c^{2}(t) dz^{2},
\end{equation}
where $a,b$ and $c,$ the directional scale factors,  are functions of time.
A special emphasis will be put on  the Kasner spacetime~\cite{kasner1921geometrical,stephani2009exact,Vish} 
\begin{equation}
 ds^{2} = -dt^{2} + t^{2p_{1}} dx^2 + t^{2p_{2}} dy^{2} + t^{2 p_{3}} dz^{2},
 \label{kas1}
\end{equation}
where the parameters $p_{1}, p_{2}$ and $p_{3}$ satisfy 
\begin{equation}
 p_{1} + p_{2} + p_{3} = p_{1}^{2} + p_{2}^{2}+p_{3}^{2} =1.
 \label{kas2}
\end{equation}
These conditions define a Kasner plane and a Kasner sphere (see 
Fig~\ref{fK3}). The Kasner metric is a solution of the vacuum Einstein field 
equations, and,  because of its simplicity, it is also a solution of the 
equations of the  quadratic  gravity. The Kasner conditions exclude the 
possibility that all three exponents are equal, however, there are 
configurations for which two of them are the same. We shall call these
configurations degenerate. 
The Kasner solution with $p_{1} = -1/3$ and $p_{2}=p_{3}=2/3$ has rotational symmetry.
The choice of $p_{i}$ in the form  $(1,0,0)$ defines the flat Kasner metric
In the nondegenerate case one can order the parameters 
$p_{i}$ as follows
\begin{equation}
 -\frac{1}{3} < p_{1} <0 <p_{2} < 2/3 < p_{3} <1 
\end{equation}
and the comoving volume element expands in two directions and compresses in one. 
In what follows however,  we shall take $p_{1}$ as an independent parameter and 
express all remaining quantities in terms of it. It can be done easily since
\begin{equation}
 p_{3} = 1-p_{1} -p_{2}
 \label{eq_p3}
\end{equation}
and 
\begin{equation}
 p_{2} = \frac{1}{2}\left(1-p_{1} -\sqrt{1 + 2 p_{1} -3 p_{1}^{2}} \right)
\end{equation}
for the lower branch, and
\begin{equation}
 p_{2} = \frac{1}{2}\left(1-p_{1} +\sqrt{1 + 2 p_{1} -3 p_{1}^{2}} \right)
\end{equation}
for the upper branch. This terminology is self-explanatory if we consider the parameter
$p_{2}$ as a function of $p_{1}$ (see Fig.~\ref{fK2}). (For different parametrizations
see Ref.~\cite{Kofman:2011tr}).
There is an interesting relation between the Kasner metric and the metric that describes 
the closest vicinity of the Schwarzschild central singularity. Indeed, it can be 
shown
that the Schwarzschild interior approaches the Kasner metric with, say, $p_{1} = 
-1/3$
and $p_{2} = p_{3}= 2/3$ as $r$ goes to 0 (see e.g. Ref.~\cite{hiscock1997,insideJM,jmP}).

The physical content of the quantum field theory in curved background is encoded 
in the regularized stress-energy tensor, $\langle T_{a}^{b}\rangle,$ and to 
certain extend in the field fluctuation,  $\langle \phi^{2} \rangle.$ In this 
formulation the spacetime is treated classically whereas the fields propagating 
on it are quantized. Even with this simplified approach the semiclassical theory 
should be able to describe quite a number of interesting phenomena, such as 
vacuum polarization, particle creation and the influence of the quantized field 
upon the background geometry~\cite{parker2009,birrell1984,fulling1989,Grib}. 
Moreover, one expects that the results obtained 
within the framework of the quantum field theory in curved background remain 
accurate as long as the quantum gravity effects are negligible. 

Ideally, the 
stress-energy tensor should depend functionally on a general metric  or at least 
on a wide class of metrics and be related to the non-local one-loop effective 
action, $W_{R},$ in a standard way. Unfortunately, such calculations are very 
hard (if not impossible) in practice. Indeed, the solutions of the field 
equations are not expressible in terms of the known special functions, the 
formal products of the operator valued distributions have to be regularized and 
the (perturbative) series are divergent. All this makes the exact analytical 
calculations practically impossible and to circumvent these problems one is 
forced either to refer to the numerical methods or to make use of some 
approximations.

In this note we shall follow the latter approach and make use of the local 
one-loop effective action constructed within the framework of the 
Schwinger-DeWitt approximation~\cite{barvinsky1985,Bryce1,FZ}, (see also 
Ref.~\cite{Ver2}). In cosmology, however,  
there is another powerful approach to the problem, namely the adiabatic 
approximation~\cite{Parker1,Parker2,Parker3,Bunch1,Bunch2,Anderson, Hubl,
Kaya,frwl2013,frwl2014,Torrenti,Ghosh} 
and closely related $n$-wave method~\cite{Yakov,Beilin,Vereshkov:1977ew,Ver2}.
For the  Robertson-Walker spacetime it has been shown that regardless of the chosen 
method, the results of the calculations are identical. It has been  demonstrated 
that the Schwinger-DeWitt approach and the adiabatic method give precisely the 
same result in this context for $4\leq D \leq 8.$ (See 
~\cite{ja2014acta,ja2016acta,delRio:2014bpa}). Building on 
this we expect that a similar correspondence also appears in the 
anisotropic homogeneous cosmologies. Although we do not attempt to perform the 
full calculations of the stress-energy tensor within the framework of the 
adiabatic approach and show that the both methods yield the same result, here we 
will solve somewhat simpler problem and demonstrate that this equality holds for 
the vacuum polarization, $\langle \phi^{2} \rangle,$ of  the massive scalar 
field with arbitrary curvature coupling in the Bianchi type I  cosmology. 
Specifically, it will be shown that the leading and the next-to-leading term  of 
the approximate vacuum polarization calculated within the framework of the 
adiabatic approximation  are precisely the same as the analogous terms 
calculated with the aid of the Schwinger-DeWitt approach. Interestingly, using 
our vacuum polarization results, we will be able to calculate the trace of the  
stress-energy tensor of the conformally coupled massive scalar field.

The paper is organized as follows. In section~\ref{sec-vac-pol} we construct the leading
and the next-to-leading term of the approximate vacuum polarization of the 
massive scalar field in the Bianchi type I  spacetime. In 
section~\ref{sec-stress-tens}  the stress-energy tensor of the scalar, spinor 
and vector fields in the Kasner spacetime is  calculated and discussed, 
whereas in Sec.~\ref{sec-back} we study the back reaction of the quantized 
fields upon the background geometry.  To the best of our knowledge 
the results of Sections~\ref{sec-vac-pol}-\ref{sec-back} are essentially new.
Throughout 
the paper the natural units are 
chosen
and we follow the Misner, Thorne and Wheeler conventions.

\section{Vacuum polarization}
\label{sec-vac-pol}

In this section we will be concerned with the neutral massive scalar field 
\begin{equation}
 - \Box \phi + (m^{2} +\xi R) \phi =0,
\label{covKG}
\end{equation}
with the arbitrary curvature coupling, $\xi,$ in the anisotropic Bianchi type I  
specetime. Our main task
is to construct the vacuum polarization. 
\subsection{Adiabatic approximation}

To simplify 
calculations we will introduce a new time coordinate~\cite{Yakov,Hu:1973kq}
\begin{equation}
 \eta = \int^{t} V^{-1/3} dt',
\end{equation}
where $V=abc,$ and redefine the field putting
\begin{equation}
 f = V^{1/3} \phi.
\end{equation}
The solution of the transformed equation can be written in the form
\begin{equation}
 f = \frac{1}{(2 \pi)^{3/2}}\int d^{3}k \left[ A_{\bf k} f_{\bf k}(\eta) e^{i {\bf k} {\bf x}} 
 + A_{\bf k}^{\dagger}f_{\bf k}^{\star}(\eta) e^{-i {\bf k} {\bf x}}  \right],
\end{equation}
where $f_{\bf k}(\eta)$ satisfies
\begin{equation}
 f_{\bf k}^{''}(\eta) + \left(\Omega^{2} + Q + Q_{1} \right) f_{\bf k}(\eta) =0
\end{equation}
with
\begin{equation}
 \Omega^{2} = V^{2/3} \left(m^{2} + \frac{k_{1}^{2}}{a^{2}} + \frac{k_{2}^{2}}{b^{2}} + \frac{k_{3}^{2}}{c^{2}}  \right),
\end{equation}
\begin{equation}
 Q = \frac{1}{3}\left( \frac{1}{3}-\xi\right)\left[ \left(\frac{a'}{a} -\frac{b'}{b} \right)^{2} +
 \left(\frac{a'}{a} -\frac{c'}{c} \right)^{2} +
 \left(\frac{b'}{b} -\frac{c'}{c} \right)^{2}\right]
\end{equation}
and
\begin{equation}
 Q_{1} = 2 \left(\xi-\frac{1}{6} \right) \left(\frac{a^{''}}{a} +\frac{b^{''}}{b}+\frac{c^{''}}{c}\right).         
\end{equation}
The $A_{\bf k}$ obey the standard commutation relations
\begin{equation}
 [ A_{\bf k},A_{\bf k'}] = [ A_{\bf k}^{\dagger},A_{\bf k'}^{\dagger}] =0
\end{equation}
and
\begin{equation}
 [A_{\bf k}, A_{\bf k'}^{\dagger}] = \delta({\bf k} - {\bf k'}),
\end{equation}
provided the functions $f_{\bf k}$ satisfy the Wronskian condition
\begin{equation}
 f_{\bf k}(\eta) f_{\bf k}^{\star '}(\eta) - f_{\bf k}^{'}(\eta) f_{\bf k}^{\star }(\eta) =i.
\end{equation}
The ground state is defined by the relation
\begin{equation}
 A_{\bf k} |0\rangle  = 0    
\end{equation}
and the formal (divergent) expression for the vacuum polarization has the 
following simple form
\begin{equation}
 \langle \phi^{2} \rangle = \frac{1}{(2 \pi)^{3} V^{2/3}} \int d^{3} k | f_{\bf 
k}|^{2}.
 \label{formal}
\end{equation}
Now, we demand that
the positive frequency functions $f_{\bf k}$ can be written in the form
\begin{equation}
 f_{\bf k} = \frac{1}{(2 V^{1/3} W)^{1/2}} \exp\left( -i \int^{\eta} V^{1/3}(s) 
W(s) ds \right),
\end{equation}
where $W(\eta)$ satisfies the following differential equation
\begin{equation}
 \Omega^{2} + Q + Q_{1} -V^{2/3} W^{2}+ \frac{7 V'^{2}}{36 V^{2}}+ \frac{V'W'}{6 V W}
 + \frac{3 W'^{2}}{4 W^{2}} - \frac{V''}{6 V} - \frac{W''}{2 W} =0.
 \label{eeq}
\end{equation}
The solution of this equation can be constructed iteratively assuming that the 
functions $W(\eta)$ (we have omitted the subscript ${\bf k}$) can be expanded as
\begin{equation}
 W = \omega_{0} + \omega_{2} + \omega_{4} + ...
 \label{serr}
\end{equation}
with the zeroth-order solution given by
\begin{equation}
 \omega_{0} = \left(m^{2} + \frac{k_{1}^{2}}{a^{2}} + \frac{k_{2}^{2}}{b^{2}} 
 + \frac{k_{3}^{2}}{c^{2}} \right)^{1/2}. 
\end{equation}
Now, in order to make our calculations more systematic and transparent, 
we shall introduce a dimensionless parameter $\varepsilon$ 
that will help to determine the adiabatic order of the complicated expressions:
\begin{equation}
 \frac{d}{d\eta}\to \varepsilon \frac{d}{d\eta} \hspace{0.5cm} {\rm and}
 \hspace{0.5cm} W =\sum_{j=0} \varepsilon^{2j} \omega_{2j},
\end{equation}
and, after the substitution of (\ref{serr}) into (\ref{eeq}), 
we collect the terms with the like powers of $\varepsilon.$
It should be noted that $\omega_{2}$ is of the second adiabatic order, 
$\omega_{4}$ is of the fourth adiabatic order, and so forth. 
Solving the chain of the algebraic equations of ascending complexity and 
substituting the thus obtained $W$ into the formal expression for the
vacuum polarization (\ref{formal}) one obtains
\begin{eqnarray}
 \langle \phi^{2} \rangle& =& \frac{1}{2 (2 \pi)^{2} V}\int \frac{d^{3}k}{\omega_{0}} 
 - \frac{\varepsilon^{2}}{2 (2 \pi)^{2} V}\int d^{3}k \frac{\omega_{2}}{\omega_{0}^{2}}
 + \frac{\varepsilon^{4}}{2 (2 \pi)^{2} V}\int d^{3}k \left( \frac{\omega_{2}^{2}}{\omega_{0}^{3}}
 -\frac{\omega_{4}}{\omega_{0}^{2}} \right)\nonumber \\
 &&-\frac{\varepsilon^{6}}{2 (2 \pi)^{2} V}\int d^{3}k \left( \frac{\omega_{2}^{3}}{\omega_{0}^{4}}
 -2 \frac{\omega_{2}\omega_{4}}{\omega_{0}^{3}}+ \frac{\omega_{6}}{\omega_{0}^{2}} \right)
 + ...
\end{eqnarray}
The first integral is divergent whereas the second one contains the terms that 
are divergent. Starting from the third term in the right hand side
of the formal expression for the vacuum polarization the integrals are finite and their computation 
in the anisotropic case presents no problems. Now, following the standard 
prescription~\cite{parker2009}
we  subtract from the formal $\langle \phi^{2} \rangle$ the first two terms, i.e., 
we subtract all the terms of a given adiabatic order  if at least one of them is 
divergent.

Thus far our analysis has been  formal. Now, let us investigate when the 
adopted method can give  well defined functions $\omega_{n}.$
In order to make our analysis more precise let us introduce the directional Hubble parameters
\begin{equation}
 H_{a} = \frac{a'}{a\, V^{1/3}}, \hspace{0.5cm}  H_{b} = \frac{b'}{b\, V^{1/3}}
 \hspace{0.5cm} {\rm and} \hspace{0.5cm}  H_{c} = \frac{c'}{c\, V^{1/3}}
\end{equation}
and observe that $\omega_{0}$  is much smaller than $\omega_{2}$ 
provided $H_{i}/m \ll 1$ where $i=a,b,c.$  Moreover, in this regime one has
\begin{equation}
 \frac{a^{(n)}}{m^{n} a V^{n/3}} \sim \left(\frac{H_{a}}{m}\right)^{n},
\end{equation}
where $a^{(n)}$ denotes $n$-th derivative of $a(\eta).$ Since similar 
relations 
for the remaining directional scale factors $b$ and $c$ hold, the magnitude of 
the terms at
the $n$-th adiabatic order are therefore equal 
to 
\begin{equation}
\left(\frac{H_{a}}{m}\right)^{\alpha}\left( \frac{H_{b}}{m}\right)^{\beta} 
\left( \frac{H_{c}}{m}\right)^{\gamma},
\end{equation}
with $\alpha+\beta+\gamma = n,$ where $\alpha, \beta,$  $\gamma$ are 
nonnegative integers. One expects that in this regime the particle creation can 
be safely ignored.

Although the $\varepsilon^{4}$-term  (i.e., the leading term) looks 
innocent it leads to the quite complicated final result. 
Indeed,  integration over angles yields 723 terms of the type
\begin{equation}
 F(a,b,c;\xi) \int \frac{dp\, p^{q}}{\left( m^{2} + p^{2} \right)^{r}},
\end{equation}
where 
\begin{equation}
p^{2} = \frac{k_{1}^{2}}{a^{2}} + \frac{k_{2}^{2}}{b^{2}} + \frac{k_{3}^{2}}{c^{2}}
\end{equation}
and the functions  $F(a,b,c;\xi)$  are constructed from $a,$ $b,$ $c$ and their 
derivatives. The number of derivatives in each term  at each adiabatic order is 
constant. Finally, after integration over $p$ one obtains 135 terms. Similarly, 
the next-to-leading term (i.e. the sixth adiabatic order) consists of 761 terms.

As the vacuum polarization of the massive scalar field with the arbitrary 
coupling  in a general Bianchi type I spacetime is rather complicated we will confine 
ourselves to the minimal and conformal couplings only.\footnote{The general 
results can be obtained on request from the author.} Integrating over $p$ we 
find
\begin{equation}
  \langle \phi^{2}\rangle^{\xi =1/6} =  \langle \phi^{2}\rangle^{\xi =0}  + \Delta,
\end{equation}
where
\begin{eqnarray}
\kappa \langle \phi^{2}\rangle^{\xi=0} &=&-\frac{2 a^{(4)}}{15 a}
+\frac{2 a^{(3)} b'}{45 a b}
+\frac{2 a^{(3)} c'}{45 a c}
+\frac{13 a'' b''}{45 a b}
+\frac{8 a'' b' c'}{45 a b c}\nonumber \\
&&
-\frac{a'' b'^2}{15 a b^2}
-\frac{a''c'^2}{15 a c^2}
+\frac{4 a''^2}{9a^2}
-\frac{14 a' b' c'^2}{135 a b c^2} 
-\frac{14 a'^3 b'}{135 a^3 b} 
-\frac{2 a'^2 b'^2}{135 a^2 b^2} \nonumber \\
&&
-\frac{14 a'^3 c'}{135 a^3 c}
+\frac{10 a'^4}{27 a^4}
+\frac{4 a^{(3)} a'}{9 a^2}
+\frac{4 a' a'' c'}{45 a^2 c}
-\frac{16 a'^2 a''}{15 a^3}
+\frac{4 b' b'' c'}{45 b^2 c} + {\rm \bf cycl},
\end{eqnarray}
\begin{eqnarray}
\kappa \Delta &=&
\frac{a^{(4)}}{9 a}
-\frac{a^{(3)} b'}{27 a b}
-\frac{a^{(3)} c'}{27 a c}
-\frac{8 a'' b''}{27 a b}
-\frac{4 a'' b' c'}{27 a b c}\nonumber \\
&&
+\frac{2 a'' b'^2}{27 a b^2}
+\frac{2 a''c'^2}{27 a c^2}
-\frac{10 a''^2}{27 a^2}
+\frac{2 a'^2 b' c'}{27 a^2 b c}
+\frac{8 a'^3 b'}{81 a^3 b}
+\frac{a'^2 b'^2}{27 a^2 b^2}\nonumber \\
&&
+\frac{8 a'^3 c'}{81 a^3 c}
-\frac{25 a'^4}{81 a^4}
-\frac{10 a^{(3)} a'}{27 a^2}
-\frac{a' a'' c'}{9 a^2 c}
+\frac{8 a'^2 a''}{9 a^3}
-\frac{b'b'' c'}{9 b^2 c}+ {\rm \bf  cycl},
\end{eqnarray}
 $\kappa =32 \pi^{2} m^{2} V^{4/3}$ and  {\bf cycl} denotes the terms that 
should be added after performing cyclic transformations 
$\{a(\eta),b(\eta),c(\eta)\} \to \{b(\eta),c(\eta),a(\eta)\}\to 
\{c(\eta),a(\eta),b(\eta)\}$. The next-to-leading term is too complicated to be  
presented here. Finally observe that the thus constructed vacuum polarization  
can easily be expressed as a function of  $t$  by a simple transformation of 
the 
time coordinate. 
 
 Now, let us return to the Kasner spacetime and calculate the first two terms of 
the vacuum polarization. Because of the simplicity of the metric the result is 
independent of the coupling constant and reads
\begin{equation}
 \langle \phi^{2} \rangle = \frac{1}{180\pi^{2} m^{2} t^{4}}\left(p_{1}^{2}
 -p_{1}^{3} \right) + \frac{1}{16 \pi^{2} m^{4} t^{6}} \left( -\frac{8}{35} 
p_{1}^{2} + \frac{8}{35} p_{1}^{3}+ 
 \frac{1}{315} p_{1}^{4} -\frac{2}{315} p_{1}^{5} + \frac{1}{315} p_{1}^{6} \right).
\end{equation}
As we shall see the second term in the right-hand-side of the above equation 
(multiplied by $m^{2}$) is 
precisely the the main approximation of the trace of the stress-energy tensor of 
the conformally coupled massive field taken with the minus sign.

\subsection{Schwinger-DeWitt approximation}

As is well known the regularized vacuum polarization of the quantized massive 
scalar field in a large mass limit can be constructed within the framework of 
the Schwinger-DeWitt method. Subtracting the terms that are divergent in the 
coincidence limit of the Schwinger-DeWitt approximation to the Green function, 
the $\langle  \phi^{2}\rangle $ can be written as a series
\begin{equation}
 \langle \phi^{2}\rangle = \frac{1}{16 \pi^{2} m^{2}} \sum_{k=2} 
 \frac{a_{k}}{(m^{2})^{k-1}} (k-2)!,
\end{equation}
where $a_{k}$ are the coincidence limit of the Hadamard-DeWitt coefficients. 
The 
usual criterion for the validity of the approximation is that the Compton 
length 
associated with the massive fields is smaller that the characteristic radius of 
curvature of the spacetime. Note, that one can safely use the conditions 
$H_{i}/m \ll 1$ in this regard. The first two coefficients have the form
\begin{equation}
 a_{2} = \frac{1}{180} R_{abcd} R^{abcd} -\frac{1}{180} R_{ab}R^{ab} + 
\frac{1}{6}\left(\frac{1}{5}-\xi \right) R_{;a}^{\phantom{a} a}+ 
\frac{1}{2}\left( \frac{1}{6}-\xi  \right)^{2} R^{2},
\end{equation}
and
\begin{equation}
 a_{3} = \frac{1}{7!} b_{3}^{(0)} + \frac{1}{360} b_{3}^{(\xi)},
\end{equation}
where
\begin{eqnarray}
b_{3}^{(0)}&=&
\frac{35}{9}R^3
+17 R_{;a}^{\phantom{;\phantom{a}}}R_{\phantom{;\phantom{a}}}^{;a}
-2 
R_{ab;c}^{\phantom{a}\phantom{b}\phantom{;\phantom{c}}}R_{\phantom{a}\phantom{b}
\phantom{;\phantom{c}}}^{ab;c}
-4 
R_{ab;c}^{\phantom{a}\phantom{b}\phantom{;\phantom{c}}}R_{\phantom{a}\phantom{c}
\phantom{;\phantom{b}}}^{ac;b}
+9 
R_{abcd;e}^{\phantom{a}\phantom{b}\phantom{c}\phantom{d}\phantom{;\phantom{e}}}
R_{\phantom{a}\phantom{b}\phantom{c}\phantom{d}\phantom{;\phantom{e}}}^{abcd;e}
\nonumber \\
&&
-8 
R_{ab;c\phantom{c}}^{\phantom{a}\phantom{b}\phantom{;\phantom{c}}c}R_{\phantom{a
}\phantom{b}}^{ab}
-\frac{14}{3} RR_{ab}^{\phantom{a}\phantom{b}}R_{\phantom{a}\phantom{b}}^{ab}
+24 
R_{ab;c\phantom{b}}^{\phantom{a}\phantom{b}\phantom{;\phantom{c}}b}R_{\phantom{a
}\phantom{c}}^{ac}
-\frac{208}{9} 
R_{ab}^{\phantom{a}\phantom{b}}R_{c\phantom{a}}^{\phantom{c}a}R_{\phantom{b}
\phantom{c}}^{bc}
\nonumber \\
&&
+ \frac{64}{3} 
R_{ab}^{\phantom{a}\phantom{b}}R_{cd}^{\phantom{c}\phantom{d}}R_{\phantom{a}
\phantom{c}\phantom{b}\phantom{d}}^{acbd}
+ 
\frac{16}{3}R_{ab}^{\phantom{a}\phantom{b}}R_{cde\phantom{a}}^{\phantom{c}
\phantom{d}\phantom{e}a}R_{\phantom{b}\phantom{e}\phantom{c}\phantom{d}}^{becd}
+ \frac{80}{9} 
R_{abcd}^{\phantom{a}\phantom{b}\phantom{c}\phantom{d}}R_{e\phantom{a}f\phantom{
c}}^{\phantom{e}a\phantom{f}c}R_{\phantom{b}\phantom{e}\phantom{d}\phantom{f}}^{
bedf}
\nonumber \\
&&
+ \frac{14}{3} 
RR_{abcd}^{\phantom{a}\phantom{b}\phantom{c}\phantom{d}}R_{\phantom{a}\phantom{b
}\phantom{c}\phantom{d}}^{abcd} 
+28 RR_{;a\phantom{a}}^{\phantom{;\phantom{a}}a}
+18 R_{;a\phantom{a}b\phantom{b}}^{\phantom{;\phantom{a}}a\phantom{b}b}
+12R_{\phantom{a}\phantom{b}\phantom{c}\phantom{d};e\phantom{e}}^{abcd\phantom{
;\phantom{e}}e}R_{abcd}^{\phantom{a}\phantom{b}\phantom{c}\phantom{d}}
\nonumber \\
&&
+\frac{44}{9} 
R_{abcd}^{\phantom{a}\phantom{b}\phantom{c}\phantom{d}}R_{ef\phantom{a}\phantom{
b}}^{\phantom{e}\phantom{f}ab}R_{\phantom{c}\phantom{d}\phantom{e}\phantom{f}}^{
cdef}
\end{eqnarray}
and
\begin{eqnarray}
b_{3}^{(\xi)}&=&
 -5R^3\xi
 +30R^3\xi^2
 -60R^3\xi^3
 -12\xi R_{;a}^{\phantom{;\phantom{a}}} R_{\phantom{;\phantom{a}}}^{;a}
 +30\xi^2 R_{;a}^{\phantom{;\phantom{a}}}R_{\phantom{;\phantom{a}}}^{;a}
-22R\xi R_{;a\phantom{a}}^{\phantom{;\phantom{a}}a}
\nonumber \\
&&
-6\xi R_{;a\phantom{a}b\phantom{b}}^{\phantom{;\phantom{a}}a\phantom{b}b}
-4\xi R_{;ab}^{\phantom{;\phantom{a}}\phantom{b}}R_{\phantom{a}\phantom{b}}^{ab}
+2R\xi R_{ab}^{\phantom{a}\phantom{b}}R_{\phantom{a}\phantom{b}}^{ab}
-2R\xi R_{abcd}^{\phantom{a}\phantom{b}\phantom{c}
\phantom{d}}R_{\phantom{a}\phantom{b}\phantom{c}\phantom{d}}^{abcd}
\nonumber \\
&&
+60R\xi^2 R_{;a\phantom{a}}^{\phantom{;\phantom{a}}a}.
\end{eqnarray}
It can be shown that calculating Hadamard-DeWitt coefficients $a_2$ and $a_{3}$ 
for the anisotropic Bianchi type I spacetime described by the line element
\begin{equation}
 ds^{2} = - V^{2/3} d\eta^{2}  + a^{2}(\eta) dx^{2} + b^{2}(\eta)dy^{2} + 
c^{2}(\eta) dz^{2},
\end{equation}
one obtains for the vacuum polarization precisely the same results 
as in the adiabatic method. 
The main approximation $a_{2}/16 \pi^{2} m^{2}$ equals the fourth-order 
adiabatic term and $a_{3}/16 \pi^{2} m^{4}$ is the same as the sixth-order 
adiabatic term. This one-to-one correspondence should hold for the higher-order terms.

\subsection{Trace of the stress-energy tensor}

At first sight it seems that the calculations reported in this section have 
nothing to do with the stress-energy tensor. However, for the conformally 
coupled fields there is an interesting relation between the trace of the 
stress-energy tensor and the vacuum polarization~\cite{Anderson:1990jh}. Indeed, 
provided $\xi =1/6$ 
one has
\begin{equation}
 T_{a}^{a} = \frac{a_{2}}{16 \pi^{2} }-m^{2} \langle \phi^{2} \rangle.
 \label{slad}
\end{equation}
Since the leading behavior of the vacuum polarization is proportional to the trace 
anomaly term, i.e., $a_{2}/(4 \pi)^{2},$ the main approximation of the trace of 
the stress-energy tensor is simply the next-to-leading term of the vacuum 
polarization taken with the minus sign. Of course it equals also the minus 
sixth-order term calculated within the framework of the adiabatic approximation. 
In the next section we shall demonstrate, among other things,  that the trace of 
the stress-energy tensor calculated from the one-loop effective action is 
given precisely by (\ref{slad}).

\section{Stress-energy tensor of quantized massive fields} 
\label{sec-stress-tens}

The one-loop effective action of the quantized fields in curved spacetime is 
nonlocal and describes both particle creation and the vacuum polarization. 
However, when the mass of the field is sufficiently large, the creation of the 
real particles is suppressed and the effective action becomes local and is 
determined by the geometry. To be more precise consider a test field of the mass 
$m$  and the associated Compton length $\lambda_{C}$ in a spacetime with the 
characteristic radius of curvature $L.$ One expects that if $\lambda_{C}/L \ll 
1$ the vacuum polarization part of the effective action dominate, making the 
expansion in inverse powers of $m^{2}$ possible. Suppose that these assumptions 
are satisfied, then the effective action is given by the Schwinger-DeWitt 
expansion~\cite{Bryce1,Bryce2,Vilkovisky,Avramidi}
\begin{equation}
 W_{R} = \frac{1}{32 \pi^{2}} \sum_{n=3}^{\infty} \frac{(n-3)!}{(m^{2})^{n-2}}  
 \int d^{4}x\, \sqrt{g}\, {\rm Tr} a_{n},
 \label{schwig_dw}
\end{equation}
where $a_{n}$ are the coincidence limit of the Hadamard-DeWitt coefficients and 
${\rm Tr}$ is a supertrace operator. Inspections of Eq.~(\ref{schwig_dw}) shows 
that the main approximation requires knowledge of the fourth Hadamard-DeWitt 
coefficients $a_{3}.$ Their exact form is known for the vector, spinor and 
scalar fields, satisfying respectively 
\begin{equation}
 (\de^{a}_{b}\Box\,-\,\na_{b}\na^{a}\,-
 \,R^{a}_{b} \,-\,\de^{a}_{b}m^{2})\phi^{(1)}_{a}\,=\,0,
 \label{Vect}
\end{equation}
\begin{equation}
(\ga^{a} \na_{a}\,+\, m) \phi^{(1/2)}\,=\,0
\label{Dirac}
\end{equation}
and Eq.~(\ref{covKG}), where $\gamma^{a}$ are the Dirac matrices. 
In the main approximation, 
the effective action of the quantized scalar, spinor and vector fields, after 
discarding total divergences and expressing the final result in the basis of the 
curvature invariants,  can be written as~\cite{Avramidi}
 \begin{eqnarray}
 W^{(1)}_{ren}\,&=& \,{1\over 192 \pi^{2} m^{2}} \int d^{4}x g^{1/2}
 \left( \al^{(s)}_{1} R \Box R
+\al^{(s)}_{2} R_{a b} \Box R^{a b}
+ \al^{(s)}_{3} R^{3}
+\al^{(s)}_{4} R R_{a b} R^{a b} 
\right. \nonumber \\
&&
+\al^{(s)}_{5} R R_{a b c d} R^{a b c d}        
+\al^{(s)}_{6} R^{\pp{b} a}_{b} R^{\pp{c} b}_{c} R^{\pp{a} c}_{a}
+\al^{(s)}_{7} R_{a b} R^{c d} R_{c \pp{a} d}^{\pp{c} a \pp{d}  b}
+\al^{(s)}_{8} R_{a b} R^{a}_{\pp{a} e c d} R^{b e c d}
\nonumber \\
&&\left. 
+\al^{(s)}_{9} R_{c d}^{\pp{c} \pp{d} a b} R_{a b}^{\pp{a} \pp{b} e h}
R_{ e h}^{\pp{e} \pp{h} c d}
+\al^{(s)}_{10} R_{c ~ d}^{~ a ~ b} R_{a ~ b}^{~ e ~ s} R_{e ~ s}^{~ c ~ d}
 \right)\nonumber \\
 &&={1\over 192 \pi^{2} m^{2}} \int d^{4}x g^{1/2} \sum_{i}^{10} \al_{i} I_{i},
\label{dzialanie}
\end{eqnarray}
where  the numerical coefficients $\alpha_{i}$ are  given in a Table~\ref{table1}. 
\begin{table}[h]
\begin{tabular}{|c|c|c|c|}\hline
& s = 0 & s = 1/2 & s = 1 \\\hline\hline
$\al^{(s)}_{1} $ & $ {1\over2}\xi^{2}\,-\,{1\over 5} \xi $\,+\,${1\over
56}$ & $- {3\over 280}$ &
$- {27\over 280}$\\
 $\al^{(s)}_{2}$ & ${1\over 140}$ & ${1\over 28}$ & ${9 \over 28}$\\
 $\al^{(s)}_{3}$ &$ \left( {1\over 6} - \xi\right)^{3}$& ${1\over 864}$ &$
- {5\over 72}$\\
 $\al^{(s)}_{4}$ & $- {1\over 30}\left( {1\over 6} - \xi\right) $& $-
{1\over 180}$ & ${31\over 60}$\\

 $\al^{(s)}_{5}$ & ${1\over 30}\left( {1\over 6} - \xi\right)$ &$ -
{7\over 1440}$ &$ - {1\over 10}$\\

 $\al^{(s)}_{6}$ & $ - {8\over 945}  $& $- {25 \over 756}$ & $- {52\over
63}$\\

 $\al^{(s)}_{7}$ & ${2 \over 315}$ & $ {47\over 1260}$  & $- {19\over 105} $\\
 $\al^{(s)}_{8}$ & ${1\over 1260}$ & ${19\over 1260} $ & ${61\over  140} $\\
 $\al^{(s)}_{9}$ & ${17\over 7560}$& ${29\over 7560}$ & $- {67\over 2520}$\\
 $\al^{(s)}_{10}$ & $- {1\over 270}$ & $ - {1\over 108} $  & $ {1\over 18}$\\ \hline
 \end{tabular}
 \caption{The coefficients $\al_{i}^{(s)}$ for the massive scalar with arbitrary
curvature coupling $\xi$ , spinor,
and vector
 field}
 \label{table1}
 \end{table}
The (renormalized) stress-energy tensor can be calculated form the standard relation
\begin{equation}
T^{ab} = \frac{2}{\sqrt{g}} \frac{\delta  W_{ren}^{(1)}}{\delta g_{ab}}
\label{set}
\end{equation}
and consists of the purely geometric terms constructed from the Riemann tensor, 
its covariant derivatives and contractions. Each $I_{i},$  after variations with 
respect to the metric tensor, leads to the covariantly conserved quantity. The 
type of the quantum field enters through the spin-dependent coefficients 
$\alpha_{i}.$ The resulting stress-energy tensor is a linear combination of 
almost 100 local geometric terms. (Their actual number depends on the 
simplification strategies and identities satisfied by the Riemann tensor used 
during the calculation). The general formulas describing the stress-energy 
tensor have been given in Refs.~\cite{ja1,ja2}.

Sometimes it is more efficient to adopt a less ambitious approach and instead of 
using general formulas of Refs.~\cite{ja2} focus on the effective action for a 
given line element and make use of the Lagrange-Euler equations. Below we shall 
briefly discuss how it can be achieved for the Kasner metric. First, observe 
that we can solve a more general problem and calculate components of the 
stress-energy tensor of the quantized fields in a general Bianchi type I 
spacetime
with $g_{00} =- f(t)$. (Henceforth, for calculational convenience, we slightly 
abuse our notation and put $g_{11} =a(t),$
$g_{22} =b(t)$ and $g_{33} =c(t)$).
Since the effective action is invariant under the cyclic transformation
\begin{equation}
 \{a(t), b(t), c(t)\} \to \{b(t),c(t), a(t)\} \to \{c(t), a(t), b(t)\}
 \label{chain}
\end{equation}
it suffices to calculate $T_{0}^{0}$ and $T_{1}^{1}.$ The remaining components
can be obtained using  this chain of  transformation in $T_{1}^{1}. $
Indeed, under the action of the cyclic transformation (\ref{chain}) one has
\begin{equation}
 T_{1}^{1}\to T_{2}^{2} \to T_{3}^{3}.
\end{equation}
For the first spatial component the Lagrange-Euler equations give
\begin{equation}
 T_{1}^{1} = \frac{a}{96\pi^{2}m^{2}\sqrt{-g}} \left[ \frac{\partial\cal{L}}{\partial a} 
 + \sum_{k=1}^{n} (-1)^{k+1}\frac{d^{k}}{dt^{k}}\left( \frac{\partial \cal{L}}{\partial a^{(k)}}\right)\right],
 \label{t11}
\end{equation}
where $\cal{L}$ is the Lagrange function density, $n$ is the maximal order of 
derivatives of the function $a(t)$ and $ a^{(k)}=d^{k} a/ dt^{k}.$ The component 
$T_{0}^{0}$ can be constructed  in a similar way.

One can also start with the simplified line element with $f(t) =1$ and using  
(\ref{t11}) calculate only $T_{1}^{1}$. The remaining spatial components can be 
obtained making use of the cyclic transformation whereas the time component of 
the stress-energy tensor can be constructed from the $\nabla_{a}T_{b}^{a} =0,$ 
which in the case at hand reduces to 
\begin{equation}
 T_{0}^{0} \left(\frac{\dot{a}}{2 a} + \frac{\dot{b}}{2 a}+\frac{\dot{b}}{2 a}\right) 
 - T_{1}^{1} \frac{\dot{a}}{2 a}-T_{2}^{2} \frac{\dot{b}}{2 b} -T_{3}^{3} \frac{\dot{c}}{2 a} + \dot{T}_{0}^{0} =0.
\end{equation}
The integration constant should be put to zero at the end of the calculations. 
Since the Ricci tensor (and Ricci scalar) vanish for the Kasner metric the 
calculations can be substantially simplified from the very beginning. For 
example, the nonvanishing terms in $\delta I_{5}/\delta g_{ab}$ and   $\delta 
I_{8}/\delta g_{ab}$ come solely from from the variations of the Ricci tensor. 
Regardless of the choice of  method the obtained results must be, of course,  
the 
same.

The stress-energy tensor of the quantized massive fields in the general Bianchi type
I spacetime is very complicated and for obvious reasons it will be not presented 
here. We only remark that the  number of terms in $T_{0}^{0}$ with the 
coefficients $c_{i}$  unspecified is 1431 whereas the number of terms in the 
each spatial component of the stress-energy tensor is 1709. For $a(t) = b(t) 
=c(t)$ the result reduces to the well-known $T_{a}^{b}$ obtained in the spatially 
flat Robertson-Walker spacetime. Although the general stress-energy tensor in 
the Bianchi type I spacetime is quite complicated, the final result for the Kasner 
metric is not. Indeed, because of spatial homogeneity there are massive 
simplifications and the resulting $T_{a}^{b}$ consists of small number of  terms 
and has a simple form that can  schematically be written as follows:
\begin{equation}
 T_{a}^{(i)b} = \frac{1}{96\pi^{2} m^{2} t^{6}} {\rm 
diag}[T^{(i)}_{0},T^{(i)}_{1},T^{(i)}_{2},T^{(i)}_{3}]_{a}^{b}
 \label{struct}
\end{equation}
where each $T^{(i)}_{k}$ is a sixth-order polynomial of $p_{1}$ and $i$ refers 
either to the upper branch or the lower branch in the parameter space. Since 
there is no danger of confusion the spin index is omitted.

\subsection{Massive scalar fields}
First let us consider the quantized massive scalar field. Making the substitution 
$\{a(t), b(t), c(t)\} \to \{t^{2p_{1}},t^{2 p_{2}}, t^{2 p_{3}}\}$ in the general
formulas in the Bianchi type I spacetime, after some algebra, one obtains (for the 
lower ($l$) and upper ($u$) branches) 
\begin{equation}
 T^{(u)0}_{0} = T^{(l)0}_{0}= \dfrac{1}{96\pi^{2} m^{2} t^6}\left[  -\frac{p_{1}^6}{21}
 +\frac{2 p_{1}^5}{21}-\frac{p_{1}^4}{21}+\frac{16 p_{1}^3}{35}-\frac{16
   p_{1}^2}{35}+\left(\frac{32 p_{1}^2}{15}-\frac{32 p_{1}^3}{15}\right) \xi \right],
\end{equation}

\begin{equation}
 T^{(u)1}_{1} = T^{(l)1}_{1}=\frac{1}{96\pi^{2} m^{2} t^{6}}\left[ \frac{p_{1}^6}{105}
 +\frac{2 p_{1}^5}{35}-\frac{11 p_{1}^4}{21}-\frac{152 p_{1}^3}{105}+\frac{40
   p_{1}^2}{21}+\left(\frac{32 p_{1}^4}{15}+\frac{32 p_{1}^3}{5}
   -\frac{128 p_{1}^2}{15}\right) \xi\right],
\end{equation}

\begin{eqnarray}
 T^{(u)2}_{2} &=& T^{(l)3}_{3} =\dfrac{1}{96\pi^{2} m^{2} t^{6}}\left[ 
 \dfrac{p_{1}^6}{105}-\dfrac{2 p_{1}^5}{35}-\dfrac{4 \beta  p_{1}^4}{105}
 +\dfrac{29 p_{1}^4}{105}-\dfrac{16 \beta 
   p_{1}^3}{105}-\dfrac{244 p_{1}^3}{105}
\right.\nonumber \\
 && \left. +\dfrac{4 \beta  p_{1}^2}{21}+\dfrac{44 p_{1}^2}{21} +\xi  \left(-\dfrac{16
   p_{1}^4}{15}+\dfrac{16 \beta  p_{1}^3}{15}+\dfrac{32 p_{1}^3}{3}
   -\dfrac{16 \beta  p_{1}^2}{15}-\dfrac{48
   p_{1}^2}{5}\right)\right]
\end{eqnarray}
and
\begin{eqnarray}
 T^{(u)3}_{3} &=& T^{(l)2}_{2} =\dfrac{1}{96\pi^{2} m^{2} t^{6}}\left[ 
 \dfrac{p_{1}^6}{105}-\dfrac{2 p_{1}^5}{35}+\dfrac{4 \beta  p_{1}^4}{105}
 +\dfrac{29 p_{1}^4}{105}+\dfrac{16 \beta 
   p_{1}^3}{105}-\dfrac{244 p_{1}^3}{105}
\right.\nonumber \\
 && \left. -\dfrac{4 \beta  p_{1}^2}{21}+\dfrac{44 p_{1}^2}{21} +\xi  
\left(-\dfrac{16
   p_{1}^4}{15}-\dfrac{16 \beta  p_{1}^3}{15}+\dfrac{32 p_{1}^3}{3}
   +\dfrac{16 \beta  p_{1}^2}{15}-\dfrac{48
   p_{1}^2}{5}\right)\right],
\end{eqnarray}
where $\beta =\sqrt{1+2 p_{1} -3 p_{1}^{2}}.$

\begin{figure}
\centering
\includegraphics[width=11cm]{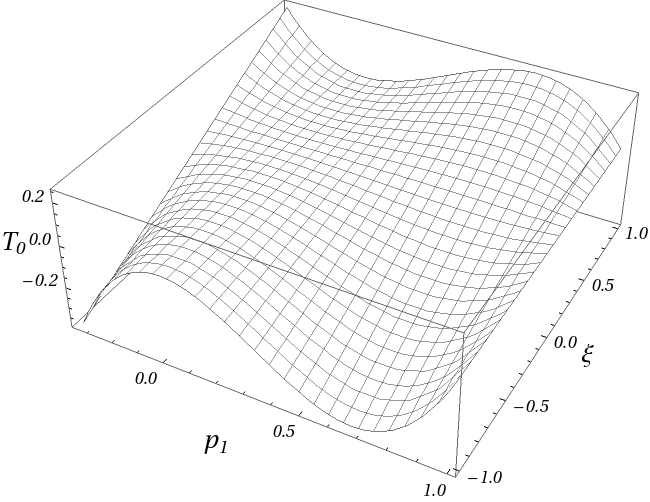}
\caption{  $T_{1} = 96 \pi^{2} m^{2} t^{6} T_{0}^{0}$ plotted as a function of $p_{1}$ and $\xi.$ }
\label{rys1}
\end{figure}

\begin{figure}
\centering
\includegraphics[width=11cm]{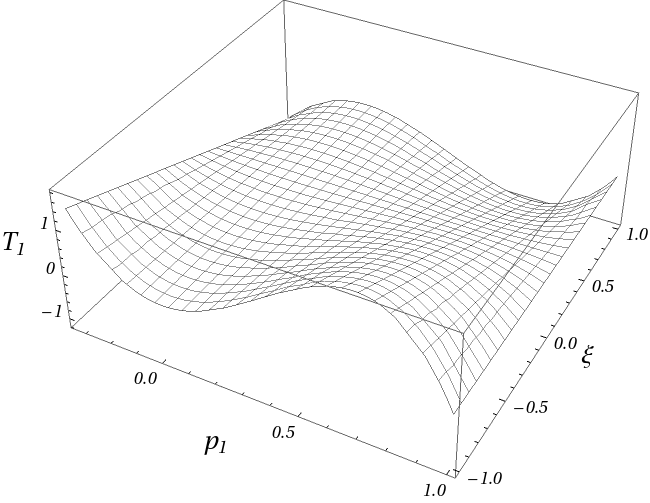}
\caption{   $T_{1} = 96 \pi^{2} m^{2} t^{6} T_{1}^{1}$ plotted as a function of $p_{1}$ and $\xi.$ }
\label{rys2}
\end{figure}

\begin{figure}
\centering
\includegraphics[width=11cm]{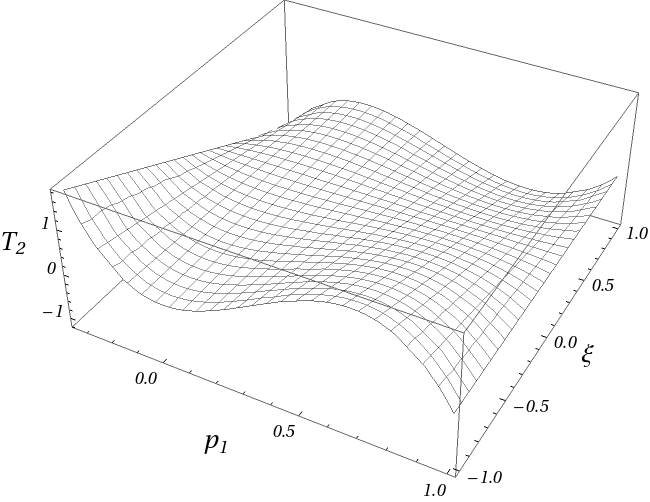}
\caption{ $T_{2} = 96 \pi^{2} m^{2} t^{6} T_{2}^{2}$ plotted as a function of $p_{1}$ and $\xi.$    }
\label{rys3}
\end{figure}

\begin{figure}
\centering
\includegraphics[width=11cm]{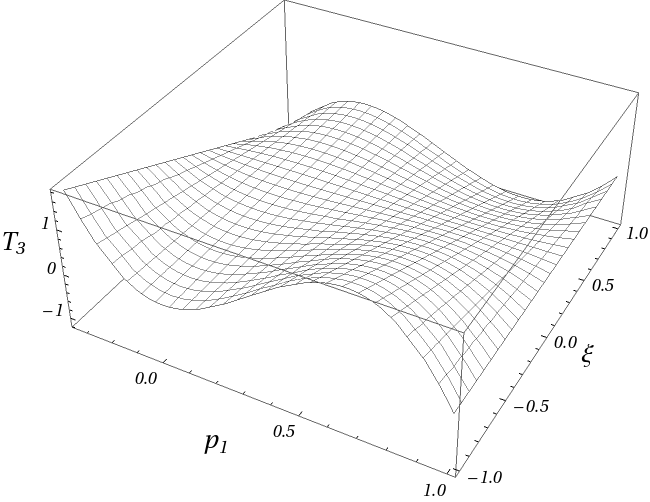}
\caption{    $T_{3} = 96 \pi^{2} m^{2} t^{6} T_{3}^{3}$ plotted as a function of $p_{1}$ and $\xi.$  }
\label{rys4}
\end{figure}

Inspection of the above formulas reveals some interesting general features: (i) 
for any allowable $p_{1}$ both $T_{0}^{0}$ and $T_{1}^{1}$ does not depend on 
the branch, (ii) the differences appear only for the remaining spatial 
components and  they are related to the change of the sign of $\beta,$ (iii) 
despite the dependence of the stress-energy tensor on the branch there are only 
four independent components as $T^{(u)2}_{2} = T^{(l)3}_{3} $ and $ T^{(u)3}_{3} 
=T^{(l)2}_{2}, $ that is in concord with the symmetries of the background 
geometry, (iv) only  $I_{5}, I_{8}, I_{9}$ and $I_{10}$  contribute to the 
final result, and, consequently, (v) the stress-energy tensor depends linearly 
on 
$\xi.$ One expects, that except (v) the features (i)-(iv) are independent of the 
spin of the quantized field. The results of the calculations for the massive scalar field 
that are plotted in 
Figs.~\ref{rys1}-\ref{rys4} reveal quite complicated (oscillatory) behavior of 
the $T_{i}$ on the $(p_{1},\xi)$-space. Here we shall focus on $\xi=0$ 
(the minimal coupling) and $\xi=1/6$ (the conformal coupling), i.e., we restrict 
ourselves to its physical values. First observe that the components of the 
stress-energy tensor are either negative or positive, and they vanish for the 
degenerate configurations for which one of the $p_{i}$ equals 1. The basic 
properties ale listed in the tables~\ref{tab1} and~\ref{tab1a}. 
\begin{center}
 \begin{table}[h]
  \begin{tabular}{|c|c|c|c|c|} \hline\hline
$\xi=0$ &    $ T_{0}^{0} \leq   0$ & $ T_{1}^{1} \geq 0$ & $ T_{2}^{2} \geq 0$ & $ T_{3}^{3} \geq 0$ \\ \hline
$\xi=1/6$ &   $ T_{0}^{0} \leq   0$ & $ T_{1}^{1} \geq 0$ & $ T_{2}^{2} \geq 0$ & $ T_{3}^{3} \geq 0$ \\ \hline\hline
  \end{tabular} 
  \caption{The sign of the components of the stress-energy tensor of the 
massive scalar field in the Kasner spacetime. The calculations have been 
carried out for the lower branch. }
  \label{tab1}
 \end{table}
\end{center}
As the functions $T_{i}$ have a simple structure 
\begin{equation}
T_{i}(p_{1})= p_{1}^{2}(p_{1}-1)  W_{3}(p_{1}),
\end{equation}
where $W_{3}(p_{1})$ is a third-order polynomial, the first local extremum of
the stress-energy tensor is always at $p_{1} =0,$ whereas location of the
second one (on the lower branch) is tabulated in Table~\ref{tab1a}.
For the upper branch the results for $T_{2}^{2}$ and $T_{3}^{3}$ should be 
interchanged. It should be noted that 
for the conformal coupling the second extremum is always at $p_{1} =2/3,$ i.e., 
for the degenerate configuration. 

\begin{center}
 \begin{table}[h]
  \begin{tabular}{|c|c|c|c|c|} \hline\hline
  & $T_{0}^{0}$ & $ T_{1}^{1}$ & $T_{2}^{2}$ & $T_{3}^{3}$\\ \hline 
$\xi=0$ &  $ p_{1} =2/3 $ &  $p_{1} = 0.6785 $ & $p_{1} =2/3 $ & $ p_{1} = 0.6547$\\ \hline\
$\xi=1/6$ &   $p_{1} =2/3 $ & $ p_{1} =2/3 $ & $ p_{1} =2/3$ & $ p_{1} =2/3$ \\ \hline\hline
  \end{tabular} 
   \caption{The extrema of the components of the stress-energy tensor of the 
massive scalar field. The calculations have been carried out for the lower 
branch. }
   \label{tab1a}
 \end{table}
\end{center}

\subsection{Massive spinor asnd vector fields}

\begin{figure}
\centering
\includegraphics[width=11cm]{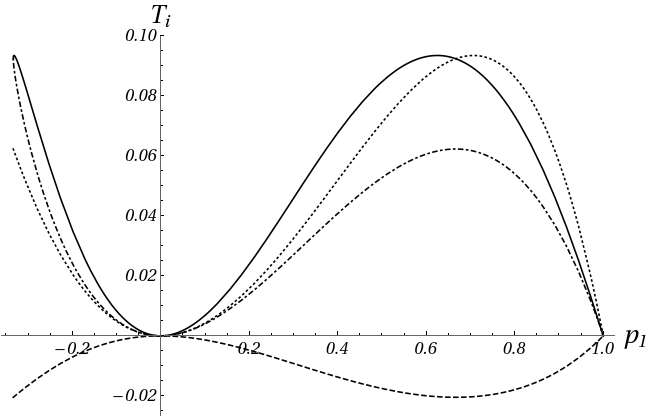}
\caption{ The functions  $T_{0}$ (dashed curve), $T_{1}$ (dotted curve), 
$T_{2}$ (dot-dashed 
curve) and $T_{3}$ (solid curve)    plotted as a function of $p_{1}.$ $T_{i}$ 
are defined as $T_{i} = 96 \pi^{2} m^{2} t^{6} T_{i}^{i}$ 
(no summation) and $s=1/2.$ }
\label{rys5}
\end{figure}

\begin{figure}
\centering
\includegraphics[width=11cm]{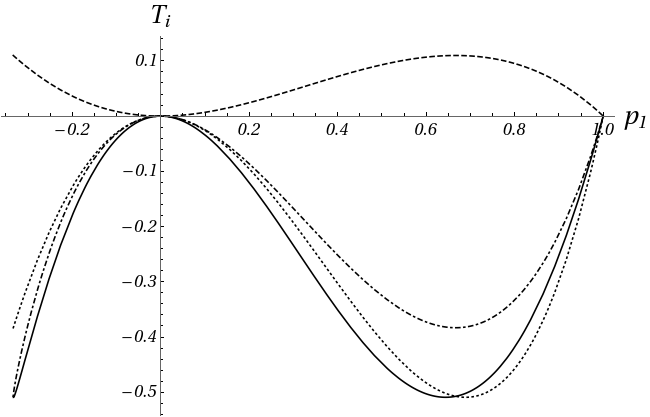}
\caption{ The functions $T_{0}$ (dashed curve), $T_{1}$ (dotted curve), $T_{2}$ 
(dot-dashed 
curve) and $T_{3}$ (solid curve)    plotted as a function of $p_{1}.$ $T_{i}$ 
are defined as $T_{i} = 96 \pi^{2} m^{2} t^{6} T_{i}^{i}$ 
(no summation) and $s =1.$   }
\label{rys6}
\end{figure}

Similar calculations can be carried out for the massive fields of higher spin. 
First, let us consider the spinor field. The stress-energy tensor in the Kasner 
spacetime has a simple form
\begin{equation}
  T^{(u)0}_{0} = \dfrac{1}{96\pi^{2} m^{2} t^6}\left(\frac{2 
p_{1}^6}{21}-\frac{4 p_{1}^5}{21}
+\frac{2 p_{1}^4}{21}+\frac{16 p_{1}^3}{105}-\frac{16 
p_{1}^2}{105}\right),
\end{equation}

\begin{equation}
 T^{(u)1}_{1} = \dfrac{1}{96\pi^{2} m^{2} t^6}\left(-\frac{2 
p_{1}^6}{105}
 -\frac{4 p_{1}^5}{35}-\frac{2 p_{1}^4}{105}-\frac{32 
p_{1}^3}{105}+\frac{16 p_{1}^2}{35}\right),
\end{equation}

\begin{equation}
 T^{(u)2}_{2} =\dfrac{1}{96\pi^{2} m^{2} t^{6}}\left(-\frac{2 
p_{1}^6}{105}
+\frac{4 p_{1}^5}{35}+\frac{8 \beta  p_{1}^4}{105}-\frac{2 p_{1}^4}{105}-\frac{8 
\beta 
   p_{1}^3}{35}-\frac{24 p_{1}^3}{35}+\frac{16 \beta  
p_{1}^2}{105}+\frac{64 p_{1}^2}{105}\right)
\end{equation}
and
\begin{equation}
 T^{(u)3}_{3} =\dfrac{1}{96\pi^{2} m^{2} t^{6}}\left( -\frac{2 
p_{1}^6}{105}+\frac{4 p_{1}^5}{35}-\frac{8 \beta 
   p_{1}^4}{105}-\frac{2 p_{1}^4}{105}+\frac{8 \beta 
   p_{1}^3}{35}-\frac{24 p_{1}^3}{35}-\frac{16 \beta 
   p_{1}^2}{105}+\frac{64 p_{1}^2}{105}\right).
\end{equation}
Similarly for the vector field one obtains
\begin{equation}
  T^{(u)0}_{0} = \dfrac{1}{96\pi^{2} m^{2} t^6}\left(-\frac{p_{1}^6}{7}
+\frac{2 p_{1}^5}{7}-\frac{p_{1}^4}{7}-\frac{16 
p_{1}^3}{21}+\frac{16 p_{1}^2}{21}\right),
\end{equation}

\begin{equation}
 T^{(u)1}_{1} = \dfrac{1}{96\pi^{2} m^{2} t^6}\left(\frac{p_{1}^6}{35}
+\frac{6 p_{1}^5}{35}+\frac{59 p_{1}^4}{105}+\frac{72 
p_{1}^3}{35}-\frac{296 p_{1}^2}{105}\right),
\end{equation}

\begin{equation}
\ T^{(u)2}_{2}  =\dfrac{1}{96\pi^{2} m^{2} t^{6}}\left(\frac{p_{1}^6}{35}
-\frac{6 p_{1}^5}{35}-\frac{4 \beta  p_{1}^4}{35}-\frac{5 p_{1}^4}{21}+\frac{64 
\beta 
   p_{1}^3}{105}+\frac{388 p_{1}^3}{105}
-\frac{52 \beta  p_{1}^2}{105}-\frac{116 p_{1}^2}{35}\right)
\end{equation}
and
\begin{equation}
 T^{(u)3}_{3}  =\dfrac{1}{96\pi^{2} m^{2} t^{6}}\left(\frac{p_{1}^6}{35}
-\frac{6 p_{1}^5}{35}+\frac{4 \beta  p_{1}^4}{35}
-\frac{5 p_{1}^4}{21}-\frac{64 \beta 
   p_{1}^3}{105}+\frac{388 p_{1}^3}{105}
+\frac{52 \beta  p_{1}^2}{105}-\frac{116 p_{1}^2}{35}\right).
\end{equation}
The results for the lower branch can be obtained, as before, form the conditions
$ T^{(l)0}_{0} = T^{(u)0}_{0},$
$T^{(l)1}_{1}=T^{(u)1}_{1},$
$T^{(l)2}_{2} =  T^{(u)3}_{3}$
and $ T^{(l)3}_{3} =  T^{(u)3}_{3}.$
The components of $T_{a}^{b}$ are still of the form given by Eq.~(\ref{struct}) 
and the functions $T_{i}$ are  plotted  in Figs.~\ref{rys5} and 
\ref{rys6}.  A comparison of the results shows that the components of the 
stress-energy tensor change their sign with a change of  spin and vanish
for the degenerate configurations of the type $(1,0,0).$ The energy density
$\rho = - T_{0}^{0}$ is nonnegative for the spinor field whereas it is negative 
(or zero) for the vector field. Moreover, the quantum effects are more 
pronounced for the spinor field. A more detailed analysis shows that
the stress-energy tensor has two local extrema: one of them is at 
$p_{1}=0$ and the location of the other is given in Table ~\ref{tab1c}.
Finally observe that for the lower branch at $p_{1} = 2/3$  the components 
$T_{1}^{(l)1}$ and $T_{3}^{(l)3}$ are equal; similarly for the upper branch 
one has $ T_{1}^{(u)1}  = T_{2}^{(u)2} $. On the other hand, at 
$p_{1} =-1/3$, (the left branch point) $ T_{2}^{2}  = T_{3}^{3}. $
This behavior is, of course, expected as  we have 
$(2/3,-1/3,2/3)$-configuration in the first case, $(2/3,2/3,-1/3)$ in the 
second, and
$(-1/3,2/3,2/3)$ in the third.

\begin{center}
 \begin{table}
  \begin{tabular}{|c|c|c|c|c|} \hline\hline
  & $T_{0}^{0}$ & $ T_{1}^{1}$ & $T_{2}^{2}$ & $T_{3}^{3}$\\ \hline 
$s = 1/2$ &  $ p_{1} =2/3 $ &  $p_{1} = 0.7067 $ & $p_{1} =2/3 $ & $ p_{1} = 
0.6250$\\ \hline\
$s = 1$ &   $p_{1} =2/3 $ & $ p_{1} =0.6890 $ & $ p_{1} =2/3$ & $ p_{1} =0.6438 
$ \\ \hline\hline
  \end{tabular} 
   \caption{The extrema of the components of the stress-energy tensor of the 
massive spinor and vector fields. The calculations have been carried out for 
the lower 
branch}
   \label{tab1c}
 \end{table}
\end{center}
\section{Back reaction on the metric}
\label{sec-back}

Although the stress-energy tensor is interesting in its own right it has much 
wider applications. Most importantly, it can be regarded as the source term of 
the semiclassical Einstein field equations
\begin{equation}
 R_{ab} - \frac{1}{2}R g_{ab} + \Lambda g_{ab} =
 8 \pi \left(T_{a}^{({ \rm cl})b} + T_{a}^{b}\right),
\end{equation}
where $T_{a}^{({ \rm cl})b}$  is the classical part of the 
total stress-energy tensor. The resulting system of differential equations has 
to be solved self-consistently for the 
quantum-corrected metric. To simplify our discussion we shall assume that the 
cosmological constant and 
the coupling parameters  $k_{1}$ and $k_{2}$ in the quadratic part of the
total action functional 
\begin{equation}
 \int d^{4} x\sqrt{g} \left(k_{1} R_{ab}R^{ab} + k_{2} R^{2}\right)
\end{equation}
vanish after the renormalization.
Unfortunately, because of the technical complexity 
of the problem it is practically impossible to find the solution of the 
equations without referring to approximations or numerics.
The exact self-consistent solutions exist only for simple geometries 
with a high degree of symmetry. Moreover, for the stress-energy tensor obtained 
from the effective action (\ref{dzialanie}) there is a real danger 
that some classes of solution of the semiclassical equations would be 
non-physical. It is because of the appearance of the higher-order derivatives 
in the  equations. Because of that our  strategy (that is in concord with the 
philosophy of the effective lagrangians) is as follows. Since the modifications 
of the classical spacetime caused by the quantum effects are expected to be 
small, the natural approach to the problem is to solve the semiclassical 
equations perturbatively.  If both the classical and the quantum parts of the 
total stress-energy tensor depend functionally on the metric, the equations to 
be solved have the following form
\begin{equation}
G_{ab}[g] =8 \pi \left(T_{a}^{({ \rm cl})b}[g] + 
\varepsilon T_{a}^{b}[g]\right),
\end{equation}
with
\begin{equation}
 g_{ab} = g^{(0)}_{ab} + \varepsilon \Delta g_{ab},
\end{equation}
where $\Delta g_{ab}$
is a first-order correction to  the metric  and  to keep control 
of the order of terms in complicated
series expansions, we have introduced once again the dimensionless parameter $\varepsilon.$ 
Focusing on the first two term of the expansion one has
\begin{equation}
 G_{ab} = G_{ab}^{(0)} + \varepsilon \Delta G_{ab}.
\end{equation}
Of course,
one expects that the quantized fields acting upon the classical Kasner   
spacetime deform it, i.e., the quantum corrected metric is still of the 
Bianchi type I type, but
it is not the Kasner metric any more (See however Ref.~\cite{Halpern}).
Having this in mind we assume that each metric potential $a(t), b(t)$ and 
$c(t)$ can be expand as the classical background plus a correction. Since 
we are interested in the corrections to the classical vacuum solution we put 
$T_{a}^{({ \rm 
cl})b}=0,$ and, consequently, 
the metric, with a little prescience, can be expanded as
\begin{eqnarray}
 a(t) = t^{2 p_{1}}\left(1 + \varepsilon \psi_{1}(t) \right),\\
 b(t) = t^{2 p_{2}}\left(1 + \varepsilon \psi_{2}(t) \right),\\
c(t) = t^{2 p_{3}}\left(1 + \varepsilon \psi_{3}(t) \right).
\end{eqnarray}
Now, expanding the semiclassical Einstein field equations in the powers of 
$\varepsilon$ and retaining the first two terms in the Einstein tensor, one has
\begin{equation}
 G^{0}_{0} = -\frac{1}{t^{2}}\left(p_{1} p_{2} + p_{1}p_{3} + p_{2}p_{3} \right)
-\frac{\varepsilon}{2t} \left(p_{2} + p_{3} \right)\psi_{1}^{\prime} 
-\frac{\varepsilon}{2t} \left(p_{1} + p_{3} \right)\psi_{2}^{\prime}
-\frac{\varepsilon}{2t} \left(p_{1} + p_{2} \right)\psi_{3}^{\prime},
\label{ee0}
\end{equation}
\begin{equation}
 G^{1}_{1} = \frac{1}{t^2}\left( p_{2} - p_{2}^{2} + p_{3}-p_{3}^2 
-p_{2}p_{3}\right)
-\frac{\varepsilon}{2t} \left(2 p_{2} + p_{3} \right)\psi_{2}^{\prime} 
-\frac{\varepsilon}{2t} \left(p_{2} + 2 p_{3} \right)\psi_{3}^{\prime}
-\frac{\varepsilon}{2}\psi_{2}^{\prime\prime}
-\frac{\varepsilon}{2}\psi_{3}^{\prime\prime},
\end{equation}
\begin{equation}
 G^{2}_{2} = \frac{1}{t^2}\left( p_{1} - p_{1}^{2} + p_{3}-p_{3}^2 
-p_{1}p_{3}\right)
-\frac{\varepsilon}{2t} \left(2 p_{1} + p_{3} \right)\psi_{1}^{\prime} 
-\frac{\varepsilon}{2t} \left(p_{1} + 2 p_{3} \right)\psi_{3}^{\prime}
-\frac{\varepsilon}{2}\psi_{1}^{\prime\prime}
-\frac{\varepsilon}{2}\psi_{3}^{\prime\prime},
\end{equation}
\begin{equation}
 G^{3}_{3} = \frac{1}{t^2}\left( p_{1} - p_{1}^{2} + p_{2}-p_{2}^2 
-p_{1}p_{2}\right)
-\frac{\varepsilon}{2t} \left(2 p_{1} + p_{2} \right)\psi_{1}^{\prime} 
-\frac{\varepsilon}{2t} \left(p_{1} + 2 p_{2} \right)\psi_{2}^{\prime}
-\frac{\varepsilon}{2}\psi_{1}^{\prime\prime}
-\frac{\varepsilon}{2}\psi_{2}^{\prime\prime}.
\end{equation}
The solution of the zeroth-order equations is the Kasner metric,  whereas  
the system of first the order  equations 
\begin{equation}
 \Delta G_{ab}=8\pi T_{a}^{b} ,
\end{equation}
where $ \Delta G_{a}^{b}$ is given by the linear in $\varepsilon $ part 
of $G_{a}^{b},$ 
is  more complicated. However, before going further it is worthwhile to
briefly discuss our general strategy. Following Ref.~\cite{Lukash:1976kr} let us assume that 
for $t<t_{0}$ $(t_{0}\gg t_{Pl})$ the stress-energy of the quantum fields vanishes. 
The modes with the frequencies satisfying  $\tilde{\omega}_{k}(t_{0}) > t_{0}^{-1}$ are in the adiabatic regime
for $t>t_{0}$ and the creation of the particles is exponentially damped. On the 
other hand, for the modes satisfying  $\tilde{\omega}_{k}(t_{0}) < t_{0}^{-1}$  we have creation,
although it can be made small taking sufficiently massive fields. In what follows 
the particle creation will be ignored.

Now return to the first order equations and analyze the degenerate configuration $(-1/3,2/3,2/3).$  
The equations to be solved are of 
the form
\begin{equation}
 \Delta G_{a}^{a} = \frac{1}{12 \pi m^{2} t^{6}} T_{a}^{(i)}\hspace{0.5cm}{\rm 
(no\, \,summation\,\, over}\, \,a).
\end{equation}
The general solution $(\psi_{1}(t),\psi_{2}(t))$ depends on three integration constants. 
The fourth one must  be equated to zero on the account of the covariant conservation of the stress-energy 
tensor. 
Since  $T_{a}^{b}(t) =0$ for $t\leq t_{0}$  we have the Kasner metric tensor and its derivative  (a left-hand
derivative at $t_{0}$) in that region.
Consequently
one is left with a simple solution
\begin{equation}
 \psi_{1}(t) = \left( \frac{1}{15}T_{1} -\frac{1}{10} T_{2}  \right)\frac{1}{t^{4}}
\end{equation}
and
\begin{equation}
 \psi_{2}(t) = - \frac{T_{1} }{12 t^{4}}
\end{equation}
with $\psi_{2}(t) = \psi_{3}(t).$
On the other hand, for a general configuration one has 
\begin{equation}
 \psi_{i} =\frac{B_{i}}{t^{4}},
\end{equation}
where $B_{i}$ for the upper branch have the form
\begin{eqnarray}
 B_{1} &=& \frac{T_{1}\left(3 p_{1}^2-2 p_{1}+15\right)}{2880 \pi 
   m^2}
  +\frac{T_{2} \left[3 (\beta  - 9)- 3 p_{1}^2+(3 \beta  -10)
   p_{1}\right]}{5760 \pi  m^2}\nonumber \\
   &&
   -\frac{T_{3} \left[3 (\beta + 9)+3
   p_{1}^2+(3 \beta +10) p_{1}\right]}{5760 \pi  m^2},
\end{eqnarray}
\begin{eqnarray}
B_{2} &=&\frac{T_{1} \left[
p_{1}(14-3p_{1}+3\beta)-5(7+\beta)
\right]}{5760 \pi  m^2}   
   +\frac{T_{2} \left[
   31+p_{1}(2-3p_{1}-3\beta)+\beta
  \right]}{5760 \pi 
   m^2}\nonumber \\
   &&
   +\frac{T_{3} \left[
   p_{1}(3 p_{1}-2)
   -4(4+\beta)
  \right]}{2880 \pi  m^2}
\end{eqnarray}
and
\begin{eqnarray}
B_{3}&=&\frac{T_{1} \left[
p_{1}(14-3p_{1}-3\beta) +5(\beta-7)
\right]}{5760 \pi  m^2}-\frac{T_{2} \left[
p_{1}(3 p_{1}-2) +4 (\beta-4)
\right]}{2880 \pi  m^2}\nonumber \\
   &&
   -\frac{T_{3}
   \left[31-\beta +p1(2-3p_{1}+3 \beta)\right]}{5760
   \pi  m^2}.
\end{eqnarray}
For a given spin,  $T_{i}$ are defined as  in Eqs.~(\ref{struct}).
The results for the lower branch can be obtained by putting $\beta \to -\beta$
and taking $T_{i}$ appropriate for that branch. With a little effort one 
can check  that the functions $\psi_{1}, \psi_{2}$ and $\psi_{3}$ satisfy 
Eq~(\ref{ee0}) and 
this completes the solution of the first order semiclassical equations.

Now we try to answer the natural question if the quantum effects dampen or 
strengthen 
the anisotropy~\cite{Lukash:1976kr}. As its  natural measure let us take the 
ratios of the 
directional Hubble parameters of the quantum-corrected spacetime. To the first 
order in $\varepsilon$ one has
\begin{equation}
 H_{ab} = \frac{H_{a}}{H_{b}} =\frac{p_{1}}{p_{2}} 
 +\frac{\varepsilon t}{2}\left(\frac{\dot{\psi}_{1}}{p_{2}}- \frac{p_{1} 
\dot{\psi}_{2}}{p_{2}^{2}} \right) ,
\end{equation}
\begin{equation}
 H_{bc} = \frac{H_{b}}{H_{c}} =\frac{p_{2}}{p_{3}} 
 +\frac{\varepsilon t}{2}\left(\frac{\dot{\psi}_{2}}{p_{3}}- \frac{p_{2} 
\dot{\psi}_{3}}{p_{3}^{2}} \right),
\end{equation}
\begin{equation}
 H_{ca} = \frac{H_{c}}{H_{a}} =\frac{p_{3}}{p_{1}} 
 +\frac{\varepsilon t}{2}\left(\frac{\dot{\psi}_{3}}{p_{1}}- \frac{p_{3} 
\dot{\psi}_{1}}{p_{1}^{2}} \right).
\end{equation}
As the result has a general structure $H_{ij} = H_{ij}^{(0)} + \delta 
H_{ij}$ we shall call  $ H_{ij}^{(0)}$ the classical part and $\delta 
H_{ij}$ its correction.
First, consider the zeroth-order effects: if the $H_{ij}^{(0)}$ is positive the 
spacetime
is expanding or contracting in the both spacetime directions, moreover, if 
$H_{ij}^{(0)} =1$ then the evolution is isotropic. On the other hand, if the 
sign is negative then the spacetime is expanding in one direction and 
contracting in the other. From this one sees that the influence of the quantum 
fields depends not only on the relative signs of the classical Hubble parameters and 
their  corrections, but also if $H_{ij}^{(0)}$ is bigger or smaller than 1. 

Before we discuss the general case let us analyze the degenerate configuration $(-1/3,2/3,2/3).$
For the massive scalar field the sign of the perturbation $\delta H_{ab}$ depends the coupling 
constant $\xi.$ Indeed, when $\xi < 47/216$ the perturbation is positive and the vacuum polarization 
isotropizes background spacetime.  It shold be noted that both minimally and conformally coupled fields make 
the background spacetime more isotropic.
Moreover, it is precisely the same inequality that should hold for the coupling constant
of the massive scalar field to make the interior of the Schwarzschild black hole more isotropic~\cite{hiscock1997}. 
It becomes even more interesting when we realize  that for the Schwarzschild black hole the degenerate 
Kasner metric is approached asymptotically only in the closest vicinity of the singularity.  
For the spinor field $\delta H_{ab}$ is always positive whereas for the vector fields it is 
always negative. Once again a similar behavior is observed for the quantum corrected  interior
of the Schwarzschild spacetime.

Now, let us return to the general case. We shall analyze the influence of the minimally and 
conformally coupled massive scalar fields on 
the  anisotropy. Here we describe only the minimally coupled fields since 
a similar qualitative behavior of $H_{ij}^{(0)}$ and $\delta H_{ij}$ can be observed 
for the conformal coupling.
On the lower branch (excluding configurations of the type $(0,0,1)$)
the  ratio $H_{ab}^{(0)}$ is always negative, $H_{bc}^{(0)}$ is 
positive for $p_{1} <0$ and negative for $p_{1}>0,$ and finally $H_{ca}^{(0)}$ 
is 
negative for $p_{1} <0$ and positive for $p_{1}>0.$ On the other hand, 
$\delta H_{ab}$ is positive for $p_{1}<0$ and negative for $p_{1}>0.$
Further, $\delta H_{bc}$ is always positive, whereas $\delta H_{ca}$  is 
negative for $p_{1} <2/3$ and positive for $p_{1}>2/3.$ 

A similar analysis carried out for the upper branch shows that $H_{ab}^{(0)}$ is negative
for $p_{1} <0$ and positive for $p_{1} >0,$ $H_{bc}^{(0)}$ is positive for $p_{1}<0$ and 
negative for $p_{1}>0,$ and $H_{ca}^{(0)}$ is always negative. The quantum correction 
$\delta H_{ab}$ is positive for $p_{1} <2/3$ and negative for $p_{1}>2/3,$ $\delta H_{bc}$  is
always negative, and, finally, $\delta H_{ca}$ is negative for $p_{1}<0$  and positive for $p_{1}>0.$

All this can be stated succinctly in the following way: roughly speaking, 
for the upper branch, the quantum effects tend to increase anisotropy  in 
$(x,y)$-directions  for $p_{1}< 2/3$ and decrease for $p_{1} >2/3.$ The 
anisotropy is always decreased by the vacuum polarization in $(y,z)$-directions 
and in $(x,z)$-directions the anisotropy is strengthened  for $p_{1} <0$ and 
damped for $p_{1}>0.$ 
On the other hand, for the lower branch the behavior of $\delta H_{12}$ is qualitatively
similar to $\delta H_{23}$ on the upper branch, whereas $\delta H_{23}$ is qualitatively
similar to $\delta H_{12}.$ The qualitative behavior of $H_{31}$ is identical on both branches.
Finally observe that for the corrections generated by the spinor and vector fields one has
a similar equivalence. More specifically, analysis of  $\delta H_{12}$ for the massive vectors
shows that the  anisotropy always increases, whereas that of  $\delta H_{23}$ increases for $p_{1}<0$ 
and decreases for $p_{1}>0.$  $\delta H_{31}$ leads to decreasing anisotropy for $p_{1}<0$
and to increasing for $p_{1}>0.$ The appropriate results for the massive spinors field are opposite, i.e.,
`increase' should be replaced by `decrease' and vice-versa.

\section{Final remarks}
\label{fin}

In this paper we have calculated the vacuum polarization, $\langle \phi^{2} \rangle,$ 
of the massive scalar field  in the Bianchi type I  spacetime within the framework
of the Schwinger-DeWitt method and the adiabatic approximation. It has been  demonstrated that
both methods yield the same result. We expect that a similar equality will hold
for the stress-energy tensors. Although we have verified this only for the trace of the 
stress-energy tensor of the conformally coupled scalar field, we believe that the demonstration of this equality in a general
case is conceptually easy but quite involved computationally. Building on this 
we have calculated the stress-energy tensor of the scalar, spinor and vector fields
in the Bianchi type I spacetime making use the Schwinger-DeWitt one-loop effective
action and checked the influence of the quantized fields upon the Kasner spacetime. 
The special emphasis has been put on the problem of isotropization
of the background geometry. It should be emphasized once again that being local 
the Schwinger-DeWitt technique  does not take particle creation into account.
It is therefore possible that the actual influence of the quantized fields, e.g., calculated numerically,
will be more pronounced~\cite{hiscock1997}. On the other hand however, we expect that 
if the conditions $H_{i}/m \ll 1$ hold our results should provide a reasonable approximation.
Finally observe that the semiclassical Einstein equations with the right hand side given by
the stress-energy tensor of the quantized fields constructed from the one-loop effective
action (\ref{dzialanie}) may be treated as the  theory with higher curvature terms. Theories of this type
are currently actively investigated (see e.g. Refs.~\cite{Pavluchenko:2017qne,Muller:2017nxg,Toporensky:2016kss} 
and references therein).

\begin{acknowledgments}
 The author would like to thank Darek Tryniecki for discussions.
\end{acknowledgments}


\begin{thebibliography}{45}
\expandafter\ifx\csname natexlab\endcsname\relax\def\natexlab#1{#1}\fi
\expandafter\ifx\csname bibnamefont\endcsname\relax
  \def\bibnamefont#1{#1}\fi
\expandafter\ifx\csname bibfnamefont\endcsname\relax
  \def\bibfnamefont#1{#1}\fi
\expandafter\ifx\csname citenamefont\endcsname\relax
  \def\citenamefont#1{#1}\fi
\expandafter\ifx\csname url\endcsname\relax
  \def\url#1{\texttt{#1}}\fi
\expandafter\ifx\csname urlprefix\endcsname\relax\def\urlprefix{URL }\fi
\providecommand{\bibinfo}[2]{#2}
\providecommand{\eprint}[2][]{\url{#2}}

\bibitem[{\citenamefont{Kasner}(1921)}]{kasner1921geometrical}
\bibinfo{author}{\bibfnamefont{E.}~\bibnamefont{Kasner}},
  \bibinfo{journal}{American Journal of Mathematics}
  \textbf{\bibinfo{volume}{43}}, \bibinfo{pages}{217} (\bibinfo{year}{1921}).

\bibitem[{\citenamefont{Stephani et~al.}(2009)\citenamefont{Stephani, Kramer,
  MacCallum, Hoenselaers, and Herlt}}]{stephani2009exact}
\bibinfo{author}{\bibfnamefont{H.}~\bibnamefont{Stephani}},
  \bibinfo{author}{\bibfnamefont{D.}~\bibnamefont{Kramer}},
  \bibinfo{author}{\bibfnamefont{M.}~\bibnamefont{MacCallum}},
  \bibinfo{author}{\bibfnamefont{C.}~\bibnamefont{Hoenselaers}},
  \bibnamefont{and} \bibinfo{author}{\bibfnamefont{E.}~\bibnamefont{Herlt}},
  \emph{\bibinfo{title}{Exact solutions of Einstein's field equations}}
  (\bibinfo{publisher}{Cambridge University Press}, \bibinfo{year}{2009}).

  \bibitem[{\citenamefont{Vish}(1984)\citenamefont{Dhurandhar,
  Vishveshwara, and Cohen}}]{Vish}
\bibinfo{author}{\bibfnamefont{S.~V.} \bibnamefont{Dhurandhar}},
  \bibinfo{author}{\bibfnamefont{C.~V.} \bibnamefont{Vishveshwara}},
  \bibnamefont{and} \bibinfo{author}{\bibfnamefont{J.~M.} \bibnamefont{Cohen}},
  \bibinfo{journal}{Class. Quant. Grav.} \textbf{\bibinfo{volume}{1}},
  \bibinfo{pages}{61} (\bibinfo{year}{1984}).




\bibitem[{\citenamefont{Kofman et~al.}(2011)\citenamefont{Kofman, Uzan, and
  Pitrou}}]{Kofman:2011tr}
\bibinfo{author}{\bibfnamefont{L.}~\bibnamefont{Kofman}},
  \bibinfo{author}{\bibfnamefont{J.-P.} \bibnamefont{Uzan}}, \bibnamefont{and}
  \bibinfo{author}{\bibfnamefont{C.}~\bibnamefont{Pitrou}},
  \bibinfo{journal}{JCAP} \textbf{\bibinfo{volume}{1105}}, \bibinfo{pages}{011}
  (\bibinfo{year}{2011}).

\bibitem[{\citenamefont{Hiscock et~al.}(1997)\citenamefont{Hiscock, Larson, and
  Anderson}}]{hiscock1997}
\bibinfo{author}{\bibfnamefont{W.~A.} \bibnamefont{Hiscock}},
  \bibinfo{author}{\bibfnamefont{S.~L.} \bibnamefont{Larson}},
  \bibnamefont{and} \bibinfo{author}{\bibfnamefont{P.~R.}
  \bibnamefont{Anderson}}, \bibinfo{journal}{Phys. Rev.}
  \textbf{\bibinfo{volume}{D56}}, \bibinfo{pages}{3571} (\bibinfo{year}{1997}).

\bibitem[{\citenamefont{Matyjasek}(2016)}]{insideJM}
\bibinfo{author}{\bibfnamefont{J.}~\bibnamefont{Matyjasek}},
  \bibinfo{journal}{Phys. Rev.} \textbf{\bibinfo{volume}{D94}},
  \bibinfo{pages}{084048} (\bibinfo{year}{2016}).

\bibitem[{\citenamefont{Matyjasek et~al.}(2013)\citenamefont{Matyjasek,
  Sadurski, and Tryniecki}}]{jmP}
\bibinfo{author}{\bibfnamefont{J.}~\bibnamefont{Matyjasek}},
  \bibinfo{author}{\bibfnamefont{P.}~\bibnamefont{Sadurski}}, \bibnamefont{and}
  \bibinfo{author}{\bibfnamefont{D.}~\bibnamefont{Tryniecki}},
  \bibinfo{journal}{Phys. Rev.} \textbf{\bibinfo{volume}{D87}},
  \bibinfo{pages}{124025} (\bibinfo{year}{2013}).

\bibitem[{\citenamefont{Parker and Toms}(2009)}]{parker2009}
\bibinfo{author}{\bibfnamefont{L.}~\bibnamefont{Parker}} \bibnamefont{and}
  \bibinfo{author}{\bibfnamefont{D.}~\bibnamefont{Toms}},
  \emph{\bibinfo{title}{Quantum field theory in curved spacetime: quantized
  fields and gravity}} (\bibinfo{publisher}{Cambridge University Press},
  \bibinfo{year}{2009}).

\bibitem[{\citenamefont{Birrell et~al.}(1984)\citenamefont{Birrell, Birrell,
  and Davies}}]{birrell1984}
\bibinfo{author}{\bibfnamefont{N.~D.} \bibnamefont{Birrell}},
  \bibinfo{author}{\bibfnamefont{N.~D.} \bibnamefont{Birrell}},
  \bibnamefont{and} \bibinfo{author}{\bibfnamefont{P.}~\bibnamefont{Davies}},
  \emph{\bibinfo{title}{Quantum fields in curved space}},
  (\bibinfo{publisher}{Cambridge University Press}, \bibinfo{year}{1984}).

\bibitem[{\citenamefont{Fulling}(1989)}]{fulling1989}
\bibinfo{author}{\bibfnamefont{S.~A.} \bibnamefont{Fulling}},
  \emph{\bibinfo{title}{Aspects of quantum field theory in curved spacetime}},
  (\bibinfo{publisher}{Cambridge University Press},
  \bibinfo{year}{1989}).

\bibitem[{\citenamefont{Grib et~al.}(1988)\citenamefont{Grib, Mamayev, and
  Mostepanenko}}]{Grib}
\bibinfo{author}{\bibfnamefont{A.~A.} \bibnamefont{Grib}},
  \bibinfo{author}{\bibfnamefont{S.~G.} \bibnamefont{Mamayev}},
  \bibnamefont{and} \bibinfo{author}{\bibfnamefont{V.~M.}
  \bibnamefont{Mostepanenko}}, \emph{\bibinfo{title}{Vacuum Quantum Effects in
  Strong Fields}} (\bibinfo{publisher}{Energoatomizdat},
  \bibinfo{address}{Moscow}, \bibinfo{year}{1988}), \bibinfo{edition}{(in
  Russian)}.

\bibitem[{\citenamefont{Barvinsky and
  Vilkovisky}(1985{\natexlab{a}})}]{barvinsky1985}
\bibinfo{author}{\bibfnamefont{A.}~\bibnamefont{Barvinsky}} \bibnamefont{and}
  \bibinfo{author}{\bibfnamefont{G.}~\bibnamefont{Vilkovisky}},
  \bibinfo{journal}{Physics Reports} \textbf{\bibinfo{volume}{119}},
  \bibinfo{pages}{1} (\bibinfo{year}{1985}{\natexlab{a}}).

\bibitem[{\citenamefont{DeWitt}(1965)}]{Bryce1}
\bibinfo{author}{\bibfnamefont{B.~S.} \bibnamefont{DeWitt}},
  \emph{\bibinfo{title}{Dynamical Theory of groups and fields}}
  (\bibinfo{publisher}{Gordon and Breach}, \bibinfo{address}{New York},
  \bibinfo{year}{1965}).

\bibitem[{\citenamefont{Frolov and Zelnikov}(1984)}]{FZ}
\bibinfo{author}{\bibfnamefont{V.~P.} \bibnamefont{Frolov}} \bibnamefont{and}
  \bibinfo{author}{\bibfnamefont{A.~I.} \bibnamefont{Zelnikov}},
  \bibinfo{journal}{Phys. Rev.} \textbf{\bibinfo{volume}{D29}},
  \bibinfo{pages}{1057} (\bibinfo{year}{1984}).

\bibitem[{\citenamefont{Vereshkov et~al.}(1989)\citenamefont{Vereshkov,
  Korotun, and Poltavtsev}}]{Ver2}
\bibinfo{author}{\bibfnamefont{G.~M.} \bibnamefont{Vereshkov}},
  \bibinfo{author}{\bibfnamefont{A.~V.} \bibnamefont{Korotun}},
  \bibnamefont{and} \bibinfo{author}{\bibfnamefont{A.~N.}
  \bibnamefont{Poltavtsev}}, \bibinfo{journal}{Sov. Phys. J.}
  \textbf{\bibinfo{volume}{32}}, \bibinfo{pages}{811} (\bibinfo{year}{1989}).

\bibitem[{\citenamefont{Parker and Fulling}(1974)}]{Parker1}
\bibinfo{author}{\bibfnamefont{L.}~\bibnamefont{Parker}} \bibnamefont{and}
  \bibinfo{author}{\bibfnamefont{S.}~\bibnamefont{Fulling}},
  \bibinfo{journal}{Phys. Rev.} \textbf{\bibinfo{volume}{D9}},
  \bibinfo{pages}{341} (\bibinfo{year}{1974}).

\bibitem[{\citenamefont{Fulling et~al.}(1974)\citenamefont{Fulling, Parker, and
  Hu}}]{Parker2}
\bibinfo{author}{\bibfnamefont{S.}~\bibnamefont{Fulling}},
  \bibinfo{author}{\bibfnamefont{L.}~\bibnamefont{Parker}}, \bibnamefont{and}
  \bibinfo{author}{\bibfnamefont{B.}~\bibnamefont{Hu}},
  \bibinfo{journal}{Phys. Rev.} \textbf{\bibinfo{volume}{D10}},
  \bibinfo{pages}{3905} (\bibinfo{year}{1974}).

\bibitem[{\citenamefont{Fulling and Parker}(1974)}]{Parker3}
\bibinfo{author}{\bibfnamefont{S.}~\bibnamefont{Fulling}} \bibnamefont{and}
  \bibinfo{author}{\bibfnamefont{L.}~\bibnamefont{Parker}},
  \bibinfo{journal}{Annals Phys.} \textbf{\bibinfo{volume}{87}},
  \bibinfo{pages}{176} (\bibinfo{year}{1974}).

\bibitem[{\citenamefont{Bunch and Davies}(1978)}]{Bunch1}
\bibinfo{author}{\bibfnamefont{T.}~\bibnamefont{Bunch}} \bibnamefont{and}
  \bibinfo{author}{\bibfnamefont{P.}~\bibnamefont{Davies}},
  \bibinfo{journal}{J. Phys.} \textbf{\bibinfo{volume}{A11}},
  \bibinfo{pages}{1315} (\bibinfo{year}{1978}).

\bibitem[{\citenamefont{Bunch}(1980)}]{Bunch2}
\bibinfo{author}{\bibfnamefont{T.}~\bibnamefont{Bunch}},
  \bibinfo{journal}{J. Phys.} \textbf{\bibinfo{volume}{A13}},
  \bibinfo{pages}{1297} (\bibinfo{year}{1980}).

\bibitem[{\citenamefont{Anderson and Parker}(1987)}]{Anderson}
\bibinfo{author}{\bibfnamefont{P.~R.} \bibnamefont{Anderson}} \bibnamefont{and}
  \bibinfo{author}{\bibfnamefont{L.}~\bibnamefont{Parker}},
  \bibinfo{journal}{Phys. Rev.} \textbf{\bibinfo{volume}{D36}},
  \bibinfo{pages}{2963} (\bibinfo{year}{1987}).
  
  \bibitem[{\citenamefont{Hu}(1978)}]{Hubl}
\bibinfo{author}{\bibfnamefont{B.~L.} \bibnamefont{Hu}},
  \bibinfo{journal}{Phys. Rev.} \textbf{\bibinfo{volume}{D18}},
  \bibinfo{pages}{4460} (\bibinfo{year}{1978}).


\bibitem[{\citenamefont{Kaya and Tarman}(2011)}]{Kaya}
\bibinfo{author}{\bibfnamefont{A.}~\bibnamefont{Kaya}} \bibnamefont{and}
  \bibinfo{author}{\bibfnamefont{M.}~\bibnamefont{Tarman}},
  \bibinfo{journal}{JCAP} \textbf{\bibinfo{volume}{1104}}, \bibinfo{pages}{040}
  (\bibinfo{year}{2011}).

\bibitem[{\citenamefont{Matyjasek and Sadurski}(2013)}]{frwl2013}
\bibinfo{author}{\bibfnamefont{J.}~\bibnamefont{Matyjasek}} \bibnamefont{and}
  \bibinfo{author}{\bibfnamefont{P.}~\bibnamefont{Sadurski}},
  \bibinfo{journal}{Phys.Rev.} \textbf{\bibinfo{volume}{D88}},
  \bibinfo{pages}{104015} (\bibinfo{year}{2013}).

\bibitem[{\citenamefont{Matyjasek et~al.}(2014)\citenamefont{Matyjasek,
  Sadurski, and Telecka}}]{frwl2014}
\bibinfo{author}{\bibfnamefont{J.}~\bibnamefont{Matyjasek}},
  \bibinfo{author}{\bibfnamefont{P.}~\bibnamefont{Sadurski}}, \bibnamefont{and}
  \bibinfo{author}{\bibfnamefont{M.}~\bibnamefont{Telecka}},
  \bibinfo{journal}{Phys. Rev.} \textbf{\bibinfo{volume}{D89}},
  \bibinfo{pages}{084055} (\bibinfo{year}{2014}).

\bibitem[{\citenamefont{Torrenti}(2015)}]{Torrenti}
\bibinfo{author}{\bibfnamefont{F.}~\bibnamefont{Torrenti}},
  \bibinfo{journal}{J. Phys. Conf. Ser.} \textbf{\bibinfo{volume}{600}},
  \bibinfo{pages}{012029} (\bibinfo{year}{2015}).

\bibitem[{\citenamefont{Ghosh}(2015)}]{Ghosh}
\bibinfo{author}{\bibfnamefont{S.}~\bibnamefont{Ghosh}},
  \bibinfo{journal}{Phys. Rev.} \textbf{\bibinfo{volume}{D91}},
  \bibinfo{pages}{124075} (\bibinfo{year}{2015}).

\bibitem[{\citenamefont{Zeldovich and Starobinsky}(1972)}]{Yakov}
\bibinfo{author}{\bibfnamefont{Y.}~\bibnamefont{Zeldovich}} \bibnamefont{and}
  \bibinfo{author}{\bibfnamefont{A.~A.} \bibnamefont{Starobinsky}},
  \bibinfo{journal}{Sov. Phys. JETP} \textbf{\bibinfo{volume}{34}},
  \bibinfo{pages}{1159} (\bibinfo{year}{1972}).

\bibitem[{\citenamefont{Beilin et~al.}(1980)\citenamefont{Beilin, Vereshkov,
  Grishkan, Ivanov, Nesterenko et~al.}}]{Beilin}
\bibinfo{author}{\bibfnamefont{V.}~\bibnamefont{Beilin}},
  \bibinfo{author}{\bibfnamefont{G.}~\bibnamefont{Vereshkov}},
  \bibinfo{author}{\bibfnamefont{Y.}~\bibnamefont{Grishkan}},
  \bibinfo{author}{\bibfnamefont{N.}~\bibnamefont{Ivanov}},
  \bibinfo{author}{\bibfnamefont{V.}~\bibnamefont{Nesterenko}},
  \bibnamefont{et~al.}, \bibinfo{journal}{Sov. Phys. JETP}
  \textbf{\bibinfo{volume}{51}}, \bibinfo{pages}{1045} (\bibinfo{year}{1980}).

\bibitem[{\citenamefont{Vereshkov et~al.}(1977)\citenamefont{Vereshkov,
  Grishkan, Ivanov, Nesterenko, and Poltavtsev}}]{Vereshkov:1977ew}
\bibinfo{author}{\bibfnamefont{G.~M.} \bibnamefont{Vereshkov}},
  \bibinfo{author}{\bibfnamefont{{\relax Yu}.~S.} \bibnamefont{Grishkan}},
  \bibinfo{author}{\bibfnamefont{S.~V.} \bibnamefont{Ivanov}},
  \bibinfo{author}{\bibfnamefont{V.~A.} \bibnamefont{Nesterenko}},
  \bibnamefont{and} \bibinfo{author}{\bibfnamefont{A.~N.}
  \bibnamefont{Poltavtsev}}, \bibinfo{journal}{Zh. Eksp. Teor. Fiz.}
  \textbf{\bibinfo{volume}{73}}, \bibinfo{pages}{1985} (\bibinfo{year}{1977}).

\bibitem[{\citenamefont{Matyjasek and Sadurski}(2014)}]{ja2014acta}
\bibinfo{author}{\bibfnamefont{J.}~\bibnamefont{Matyjasek}} \bibnamefont{and}
  \bibinfo{author}{\bibfnamefont{P.}~\bibnamefont{Sadurski}},
  \bibinfo{journal}{Acta Phys. Polon.} \textbf{\bibinfo{volume}{B45}},
  \bibinfo{pages}{2027} (\bibinfo{year}{2014}).

\bibitem[{\citenamefont{Matyjasek and Tryniecki}(2016)}]{ja2016acta}
\bibinfo{author}{\bibfnamefont{J.}~\bibnamefont{Matyjasek}} \bibnamefont{and}
  \bibinfo{author}{\bibfnamefont{D.}~\bibnamefont{Tryniecki}},
  \bibinfo{journal}{Acta Phys. Polon.} \textbf{\bibinfo{volume}{B47}},
  \bibinfo{pages}{2095} (\bibinfo{year}{2016}).

\bibitem[{\citenamefont{del Rio and Navarro-Salas}(2015)}]{delRio:2014bpa}
\bibinfo{author}{\bibfnamefont{A.}~\bibnamefont{del Rio}} \bibnamefont{and}
  \bibinfo{author}{\bibfnamefont{J.}~\bibnamefont{Navarro-Salas}},
  \bibinfo{journal}{Phys. Rev.} \textbf{\bibinfo{volume}{D91}},
  \bibinfo{pages}{064031} (\bibinfo{year}{2015}).

\bibitem[{\citenamefont{Hu et~al.}(1973)\citenamefont{Hu, Fulling, and
  Parker}}]{Hu:1973kq}
\bibinfo{author}{\bibfnamefont{B.~L.} \bibnamefont{Hu}},
  \bibinfo{author}{\bibfnamefont{S.~A.} \bibnamefont{Fulling}},
  \bibnamefont{and} \bibinfo{author}{\bibfnamefont{L.}~\bibnamefont{Parker}},
  \bibinfo{journal}{Phys. Rev.} \textbf{\bibinfo{volume}{D8}},
  \bibinfo{pages}{2377} (\bibinfo{year}{1973}).

\bibitem[{\citenamefont{Anderson}(1990)}]{Anderson:1990jh}
\bibinfo{author}{\bibfnamefont{P.~R.} \bibnamefont{Anderson}},
  \bibinfo{journal}{Phys. Rev.} \textbf{\bibinfo{volume}{D41}},
  \bibinfo{pages}{1152} (\bibinfo{year}{1990}).

\bibitem[{\citenamefont{DeWitt}(1975)}]{Bryce2}
\bibinfo{author}{\bibfnamefont{B.~S.} \bibnamefont{DeWitt}},
  \bibinfo{journal}{Phys. Rept.} \textbf{\bibinfo{volume}{19}},
  \bibinfo{pages}{295} (\bibinfo{year}{1975}).

\bibitem[{\citenamefont{Barvinsky and
  Vilkovisky}(1985{\natexlab{b}})}]{Vilkovisky}
\bibinfo{author}{\bibfnamefont{A.~O.} \bibnamefont{Barvinsky}}
  \bibnamefont{and} \bibinfo{author}{\bibfnamefont{G.~A.}
  \bibnamefont{Vilkovisky}}, \bibinfo{journal}{Phys. Rept.}
  \textbf{\bibinfo{volume}{119}}, \bibinfo{pages}{1}
  (\bibinfo{year}{1985}{\natexlab{b}}).

\bibitem[{\citenamefont{Avramidi}(1989)}]{Avramidi}
\bibinfo{author}{\bibfnamefont{I.}~\bibnamefont{Avramidi}},
  \bibinfo{journal}{Theor. Math. Phys.} \textbf{\bibinfo{volume}{79}},
  \bibinfo{pages}{494} (\bibinfo{year}{1989}).

\bibitem[{\citenamefont{Matyjasek}(2000)}]{ja1}
\bibinfo{author}{\bibfnamefont{J.}~\bibnamefont{Matyjasek}},
  \bibinfo{journal}{Phys. Rev.} \textbf{\bibinfo{volume}{D61}},
  \bibinfo{pages}{124019} (\bibinfo{year}{2000}).

\bibitem[{\citenamefont{Matyjasek}(2001)}]{ja2}
\bibinfo{author}{\bibfnamefont{J.}~\bibnamefont{Matyjasek}},
  \bibinfo{journal}{Phys. Rev.} \textbf{\bibinfo{volume}{D63}},
  \bibinfo{pages}{084004} (\bibinfo{year}{2001}).

\bibitem[{\citenamefont{Halpern}(1994)}]{Halpern}
\bibinfo{author}{\bibfnamefont{P.}~\bibnamefont{Halpern}},
  \bibinfo{journal}{Gen. Rel. Grav.} \textbf{\bibinfo{volume}{26}},
  \bibinfo{pages}{781} (\bibinfo{year}{1994}).

\bibitem[{\citenamefont{Lukash et~al.}(1976)\citenamefont{Lukash, Novikov,
  Starobinsky, and Zeldovich}}]{Lukash:1976kr}
\bibinfo{author}{\bibfnamefont{V.~N.} \bibnamefont{Lukash}},
  \bibinfo{author}{\bibfnamefont{I.~D.} \bibnamefont{Novikov}},
  \bibinfo{author}{\bibfnamefont{A.~A.} \bibnamefont{Starobinsky}},
  \bibnamefont{and} \bibinfo{author}{\bibfnamefont{{\relax Ya}.~B.}
  \bibnamefont{Zeldovich}}, \bibinfo{journal}{Nuovo Cim.}
  \textbf{\bibinfo{volume}{B35}}, \bibinfo{pages}{293} (\bibinfo{year}{1976}).

\bibitem[{\citenamefont{Pavluchenko and
  Toporensky}(2018)}]{Pavluchenko:2017qne}
\bibinfo{author}{\bibfnamefont{S.~A.} \bibnamefont{Pavluchenko}}
  \bibnamefont{and}
  \bibinfo{author}{\bibfnamefont{A.}~\bibnamefont{Toporensky}},
  \bibinfo{journal}{Eur. Phys. J.} \textbf{\bibinfo{volume}{C78}},
  \bibinfo{pages}{373} (\bibinfo{year}{2018}).

\bibitem[{\citenamefont{Müller et~al.}(2018)\citenamefont{Müller,
  Ricciardone, Starobinsky, and Toporensky}}]{Muller:2017nxg}
\bibinfo{author}{\bibfnamefont{D.}~\bibnamefont{Müller}},
  \bibinfo{author}{\bibfnamefont{A.}~\bibnamefont{Ricciardone}},
  \bibinfo{author}{\bibfnamefont{A.~A.} \bibnamefont{Starobinsky}},
  \bibnamefont{and}
  \bibinfo{author}{\bibfnamefont{A.}~\bibnamefont{Toporensky}},
  \bibinfo{journal}{Eur. Phys. J.} \textbf{\bibinfo{volume}{C78}},
  \bibinfo{pages}{311} (\bibinfo{year}{2018}).

\bibitem[{\citenamefont{Toporensky and M{\"u}ller}(2017)}]{Toporensky:2016kss}
\bibinfo{author}{\bibfnamefont{A.}~\bibnamefont{Toporensky}} \bibnamefont{and}
  \bibinfo{author}{\bibfnamefont{D.}~\bibnamefont{M{\"u}ller}},
  \bibinfo{journal}{Gen. Rel. Grav.} \textbf{\bibinfo{volume}{49}},
  \bibinfo{pages}{8} (\bibinfo{year}{2017}).

\end{thebibliography}

\end{document}